\documentclass[thmsa,a4paper]{article}
\usepackage{amssymb}
\usepackage{sw20asas}

\input tcilatex
\begin{document}

\begin{center}
* * *

DIPOLE \& ABSOLUTE MAGNITUDE ANALYSIS OF THE SCP UNION\ SUPERNOVAE

WITHIN THE EXPANSION CENTER MODEL

\vspace{.1in}

ECM paper XVI by Luciano Lorenzi

by merging the SAIt 2011 ECM paper XI with the EWASS 2012 ECM paper XIII

\vspace{.15in}

ABSTRACT
\end{center}

\begin{quotation}
\textbf{1743 data calculated for 249 High-}$z$\textbf{\ SCP Union supernovae
are analysed according to the expansion center model (ECM). The analysis in
Hubble units begins with 13 listed normal points corresponding to 13 }$z$%
\textbf{-bin samples at as many Hubble depths. The novel finding is a clear
drop in the average scattering of the SNe Ia Hubble Magnitude }$M$\textbf{\
with the Hubble depth }$D$\textbf{, after using the average trend }$\langle
M\rangle $\textbf{\ computed in paper IX. Other correlations of the }$M$ 
\textbf{scattering with the position in the sky are proposed. Consequently,
13 ECM dipole tests on the 13 }$z$\textbf{-bin samples were carried out both
with unweighted and weighted fittings. A further check was made with Hubble
depths }$D$\textbf{\ obtained by assuming }$M\equiv \langle M\rangle $%
\textbf{\ according to paper IX and XV. In conclusion the analysis of 249 }$%
SCPU$\textbf{\ SNe confirms once again the ECM at any Hubble depth,
including a strengthening }$\Delta M$ \textbf{perturbation effect at
decreasing }$z\lesssim 0.5$. \textbf{A new successful dipole test introduces
the absolute magnitude analysis of 398 }$SCPU$\textbf{\ supernovae. After
testing 14 high-}$z$ \textbf{normal points }$\langle M_B\rangle $\textbf{\
from paper IX Table 2, a trend analysis of another 15 and 30 normal points
of the Hubble Magnitude }$M$ \textbf{and a new absolute magnitude }$M^{*}$, 
\textbf{at increasing }$\langle z\rangle \equiv z_0$\textbf{\ corresponding
to a different series of }$z$\textbf{\ bins, leads to the discovery of the
magnitude anomaly of the low }$\langle z\rangle $\textbf{\ points. When the
low }$\langle z\rangle $\textbf{\ points are excluded, the best fittings
make it possible to extrapolate the SNe Ia absolute magnitude }$M_0$ \textbf{%
at a central redshift }$z_0\rightarrow 0$\textbf{, with }$M_0=-17.9\pm 0.1$%
\textbf{\ and a few final ECM\ solutions of the SNe Ia }$\langle M\rangle $%
\textbf{\ and }$\langle M^{*}\rangle $\textbf{. The magnitude anomaly is
here interpreted as due to a deficiency in the magnitude formulas used;
these produce a maximum peak of deviation, with a systematic }$\Delta
M\approx 1$ \textbf{in the range }$0.04\lesssim $\textbf{\ }$\langle
z\rangle \lesssim 0.08$.\textbf{\ That is a proof of the Universe rotation
within the expansion center model.}
\end{quotation}

\newpage

\section{Introduction}

The present work, which results to be a fusion of paper XI with paper XIII
presented at EWASS 2012, is to all intents a further and necessary
supplement to complete the parallel paper XV, which represents the final
crucial proof of the expansion center Universe. In that paper, the model
independent dipole test was limited to $z$ bins centred on $\langle z\rangle
=1$. Here the aim is the ECM dipole analysis of 13 $z$ bins at different
Hubble depths, using 249 supernovae lying within the range $0.2<z<1.4$ from
the selected 307 SNe Ia of the SCP Union compilation ($SCPU$: Kowalski et.
al. 2008), in order to show how the wedge-shaped Hubble diagram of paper IX
is affected by both the ECM dipole anisotropy and a $\Delta M$ effect that
appears to be more perturbative at decreasing $z\lesssim 0.5$. Indeed,
having confirmed the expansion center model (ECM), here the ECM is used to
check and explore more thoroughly the SNe Ia behaviour at varying Hubble
depths and positions in the sky. Hence the analysis of the SNe Ia absolute
magnitude bases itself on the data of the whole SCP Union sample, which
reports redshifts and blue apparent magnitudes of 398 SNe Ia. Owing to the
cited strengthening perturbation effect of the scattering of the Hubble
Magnitude $M$ at decreasing $z<0.5$, new $\langle M\rangle $ fittings
limited to normal points with $\langle z\rangle >0.55$ from paper IX Table 2
have been explored. After the successful check, 30 new normal points from
the data of all the 398 $SCPU$ SNe have been constructed, in order to better
analyse the SN magnitude trend at different Hubble depths. The main
construction and analysis of the magnitude normal points does not involve
the expansion center model. In other words the main experimental results
obtained, the SNe absolute magnitude value $M_0$ and the trend of the Hubble
Magnitude $M$, can be considered both model independent and able to confirm
once again the ECM. In particular the new findings provide astronomical
evidence for cosmic rotation around the expansion center, in accordance with
the limits of the ECM itself, which formally, as one must recall, implies a
rigid rotation of the very nearby Universe (cf. paper VII).

All the plots and graphical fittings of this analysis appear in the Appendix
''Atlas of the ECM paper XVI figures''. Moreover, as we deal only with blue
magnitudes, the pedicel $B$ becomes superfluous; thus the convention $%
M_B\equiv M$ is adopted within this paper XVI as in paper XV.

The cited papers I-II-III-IV-V-VI-VII-VIII-IX-X-XI-XII-XIII-XV are those
referenced as Lorenzi 1999a$\rightarrow $2012d, while S\&T is for Sandage \&
Tammann 1975a, B\&S for Bahcall \& Soneira 1982, P99 for Perlmutter et al.
1999, K03 for Knop et al. 2003.

\section{ECM values from the observed ($z,m,\gamma $)}

\subsection{ECM standard values}

The ECM Hubble law in Hubble units (cf. papers V-VI-IX), 
\begin{equation}
cz=\left[ H_0-a_0X\right] \cdot D=H_X\cdot D\hspace{.4in}with\hspace{.4in}%
H_X=H_0-a_0X
\end{equation}
where 
\begin{equation}
X=\cos \gamma \cdot (1-x)^{\frac 13}/(1+x)\hspace{.3in}D=\frac{xc}{3H_0}%
\left( \frac{1+x}{1-x}\right) \hspace{.3in}r=\frac{xc}{3H_0}
\end{equation}
and after introducing the ECM standard values (based on data by S\&T) 
\begin{equation}
\mathbf{H}_0\mathbf{\equiv 70}\text{ }km\text{ }s^{-1}Mpc^{-1}\mathbf{%
\hspace{.4in}a}_0\mathbf{\equiv 12.7}\text{ }km\text{ }s^{-1}Mpc^{-1}
\end{equation}
allows us to give each supernova at $(\alpha ,\delta )$ the ECM light space $%
r_z$, Hubble depth $D_z$ and Hubble Magnitude $M_z=M(D_z)$, being $D_L\equiv
D_C$ assumed (cf. papers V-VI-IX-XV) and $z,m=m_B^{\max }$ available from
literature together with the computed value of $\cos \gamma =\sin \delta
_{VC}\sin \delta +\cos \delta _{VC}$ $\cos \delta $ $\cos (\alpha -\alpha
_{VC})$ with $\alpha _{VC}\approx 9^h$ and $\delta _{VC}\approx +30^0$
(B\&S), as follows: 
\begin{equation}
\left[ z,\cos \gamma \right] \Rightarrow x=x(z,\cos \gamma )\Rightarrow
r=r_z;\text{ }D=D_z;\text{ }X=X_z\Rightarrow cz\equiv H_X\cdot D_z\Rightarrow
\end{equation}
\begin{equation}
\left[ m=m_B^{\max }\right] \Rightarrow D_C=D_z\cdot (1+z)=\frac{xc}{3H_0}%
\left( \frac{1+x}{1-x}\right) (1+z)\Rightarrow M_z=m-5\log D_C-25
\end{equation}

\subsection{Computation of the $M$ scattering}

The ECM Hubble Magnitude $M_z$ needs to be compared with the model
independent value $\langle M\rangle $, which comes from the $M(D)$ average
trend computed in paper IX, whose eq. (22) gives the fitting curve of 30
normal points from 398 SNe listed in Table 11 of the $SCPU$ compilation
(Kowalski et al. 2008). Here the computation of $\langle M\rangle $, then
the scattering $\Delta M=M_z-\langle M\rangle $, utilizes the ECM Hubble
depth $D_z$ with the same parameters $d_0d_1d_2$, according to the following
expressions: 
\begin{equation}
\cos \gamma =0\Rightarrow z=z_0\Rightarrow D=\frac{cz_0}{H_0}\hspace{.3in}and%
\hspace{.3in}\cos \gamma \neq 0\Rightarrow D\text{ }=\frac{cz}{H_X}%
=D_z\Rightarrow
\end{equation}
\begin{equation}
M(z_0)=A_0+A_1z_0+A_2z_0^2=d_0+d_1D_z+d_2D_z^2=\langle M\rangle
\end{equation}
\begin{equation}
d_0=A_0\text{ }\cong -18.77\text{ };\text{ }d_1\cong -1.421\cdot H_0/c\text{ 
};\text{ }d_2\cong +0.3589\cdot H_0^2/c^2
\end{equation}
\begin{equation}
\Delta M=M_z-\langle M\rangle
\end{equation}

\subsection{Computation of the Hubble depth $D_{\langle M\rangle }$}

On the basis of the previous formulations, one can finally calculate the
Hubble depth $D_{\langle M\rangle }$ corresponding to the Hubble Magnitude
equal to its average value $\langle M\rangle $ of eq. (7). To this end
consider some sequential steps: 
\begin{equation}
\langle M\rangle =M_z-\Delta M=m-5\log D_C-25-\Delta M
\end{equation}
\begin{equation}
\log \left[ D_z(1+z)\right] =0.2(m-\langle M\rangle -\Delta M)-5
\end{equation}
\begin{equation}
D_z=10^{0.2(m-\langle M\rangle -\Delta M)-5}/(1+z)
\end{equation}
\begin{equation}
D_{\langle M\rangle }=10^{0.2(m-\langle M\rangle )-5}/(1+z)
\end{equation}
\begin{equation}
D_{\langle M\rangle }=D_z\cdot 10^{0.2\Delta M}
\end{equation}
\begin{equation}
\Delta M\rightarrow 0\Rightarrow M_z\rightarrow \langle M\rangle \Rightarrow
D_{\langle M\rangle }\rightarrow D_z
\end{equation}

\subsection{Weighted ECM Hubble law}

It is clear that only the Hubble depth $D_{\langle M\rangle }$ can be
obtained, if we exclude the ECM value of $D_z$. So the check of $a_0$ in eq.
(1), after introducing $D=$ $D_{\langle M\rangle }$ from eq. (13), should
take into account the $\Delta M$ perturbation. That may be done through an
ECM Hubble law weighted by $w_i$ with $i=0,1,2$, as follows: 
\begin{equation}
cz=\left( H_0-a_0X_z\right) \cdot D_{\langle M\rangle }\Rightarrow Y=\frac{cz%
}{D_{\langle M\rangle }}-H_0\rightarrow -X_z\Rightarrow a_0\hspace{.4in}with%
\hspace{.2in}w_i\propto \left| \Delta M\right| ^{-i}
\end{equation}

\section{ECM dipole analysis of 249 High-$z$ SCP Union SNe Ia}

The ECM values for each of the 249 SNe Ia (those of Table 3 in Appendix of
paper IX and cited in the papers X-XV as pilot sample XVI) are listed in the
corresponding Table 3 of section 4.

\subsection{Construction of 13 normal points}

Table 0 below lists a set of normal points referring to the pilot sample XVI
and to 12 derived samples. In particular the $10$ columns of Table 0 present
the following data for each SNe sample, in order: Sample ordinal number;
number N of the sample SNe; sample $z$ bin; mean $\langle z\rangle $ of the $%
z$ bin; unweighted mathematical mean $\langle m_B^{\max }\rangle $ of the
corresponding SNe magnitudes; mean $\langle \cos \gamma \rangle $ of the SNe 
$\cos \gamma $; relative scattering of the average Hubble depth of the
sample SNe, as $\frac{\Delta D}D$ where $\Delta D=\langle D_z\rangle -D$ and 
$D=\frac{c\langle z\rangle }{H_0}$ with $\langle z\rangle \equiv z_0$
assumed; average Hubble depth of the sample SNe, as $\langle D_z\rangle $,
whose individual $D_z$ come from the ECM solution (4); average Hubble
Magnitude $\langle M_z\rangle $ of the sample SNe whose individual $M_z$
follow from eq. (5); average scattering in modulus of the sample SNe Hubble
Magnitudes, as $\langle \left| \Delta M\right| \rangle $ whose individual $%
\left| \Delta M\right| $ $=\left| M_z-\langle M\rangle \right| $ follow from
the $M_z$ eq. (5) and the average trend $\langle M\rangle
=d_0+d_1D_z+d_2D_z^2$\ according to eqs. (7)(8).

\vspace{.05in}

\textbf{Table 0}

\vspace{.1in}

\begin{tabular}{|cccccccccc|}
\hline
\multicolumn{1}{|c|}{Sample} & \multicolumn{1}{c|}{N} & \multicolumn{1}{c|}{$%
z$ bin} & \multicolumn{1}{c|}{$\langle z\rangle $} & \multicolumn{1}{c|}{$%
\langle m\rangle $} & \multicolumn{1}{c|}{$\langle \cos \gamma \rangle $} & 
\multicolumn{1}{c|}{$\frac{\Delta D}D$} & \multicolumn{1}{c|}{$\langle
D_z\rangle $} & \multicolumn{1}{c|}{$\langle M_z\rangle $} & $\langle \left|
\Delta M\right| \rangle $ \\ \hline
\multicolumn{1}{|c|}{XVI$_{11}$} & \multicolumn{1}{c|}{50} & 
\multicolumn{1}{c|}{$0.2<z\leq 0.4$} & \multicolumn{1}{c|}{$0.322$} & 
\multicolumn{1}{c|}{22.123} & \multicolumn{1}{c|}{$-0.00$} & 
\multicolumn{1}{c|}{0.03} & \multicolumn{1}{c|}{1425} & \multicolumn{1}{c|}{$%
-19.146$} & $0.342$ \\ \hline
\multicolumn{1}{|c|}{XVI$_{12}$} & \multicolumn{1}{c|}{101} & 
\multicolumn{1}{c|}{$0.2<z\leq 0.5$} & \multicolumn{1}{c|}{$0.387$} & 
\multicolumn{1}{c|}{22.534} & \multicolumn{1}{c|}{$+0.06$} & 
\multicolumn{1}{c|}{0.04} & \multicolumn{1}{c|}{1716} & \multicolumn{1}{c|}{$%
-19.235$} & $0.330$ \\ \hline
\multicolumn{1}{|c|}{XVI$_{13}$} & \multicolumn{1}{c|}{142} & 
\multicolumn{1}{c|}{$0.2<z\leq 0.6$} & \multicolumn{1}{c|}{$0.434$} & 
\multicolumn{1}{c|}{22.762} & \multicolumn{1}{c|}{$+0.03$} & 
\multicolumn{1}{c|}{0.02} & \multicolumn{1}{c|}{1893} & \multicolumn{1}{c|}{$%
-19.304$} & $0.302$ \\ \hline
\multicolumn{1}{|c|}{XVI$_{14}$} & \multicolumn{1}{c|}{174} & 
\multicolumn{1}{c|}{$0.2<z\leq 0.7$} & \multicolumn{1}{c|}{$0.472$} & 
\multicolumn{1}{c|}{22.928} & \multicolumn{1}{c|}{$+0.00$} & 
\multicolumn{1}{c|}{0.01} & \multicolumn{1}{c|}{2037} & \multicolumn{1}{c|}{$%
-19.357$} & $0.289$ \\ \hline
\multicolumn{1}{|c|}{XVI$_{15}$} & \multicolumn{1}{c|}{192} & 
\multicolumn{1}{c|}{$0.2<z\leq 0.8$} & \multicolumn{1}{c|}{$0.499$} & 
\multicolumn{1}{c|}{23.040} & \multicolumn{1}{c|}{$+0.03$} & 
\multicolumn{1}{c|}{0.01} & \multicolumn{1}{c|}{2152} & \multicolumn{1}{c|}{$%
-19.389$} & $0.279$ \\ \hline
\multicolumn{1}{|c|}{XVI$_{16}$} & \multicolumn{1}{c|}{215} & 
\multicolumn{1}{c|}{$0.2<z\leq 0.9$} & \multicolumn{1}{c|}{$0.535$} & 
\multicolumn{1}{c|}{23.195} & \multicolumn{1}{c|}{$+0.05$} & 
\multicolumn{1}{c|}{0.02} & \multicolumn{1}{c|}{2332} & \multicolumn{1}{c|}{$%
-19.417$} & $0.271$ \\ \hline
\multicolumn{1}{|c|}{XVI} & \multicolumn{1}{c|}{249} & \multicolumn{1}{c|}{$%
0.2<z<1.4$} & \multicolumn{1}{c|}{$0.607$} & \multicolumn{1}{c|}{23.440} & 
\multicolumn{1}{c|}{$+0.08$} & \multicolumn{1}{c|}{0.02} & 
\multicolumn{1}{c|}{2651} & \multicolumn{1}{c|}{$-19.482$} & $0.259$ \\ 
\hline
\multicolumn{1}{|c|}{XVI$_{17}$} & \multicolumn{1}{c|}{200} & 
\multicolumn{1}{c|}{$0.4\leq z<1.4$} & \multicolumn{1}{c|}{$0.677$} & 
\multicolumn{1}{c|}{23.763} & \multicolumn{1}{c|}{$+0.10$} & 
\multicolumn{1}{c|}{0.02} & \multicolumn{1}{c|}{2951} & \multicolumn{1}{c|}{$%
-19.567$} & $0.239$ \\ \hline
\multicolumn{1}{|c|}{XVI$_{18}$} & \multicolumn{1}{c|}{149} & 
\multicolumn{1}{c|}{$0.5\leq z<1.4$} & \multicolumn{1}{c|}{$0.756$} & 
\multicolumn{1}{c|}{24.052} & \multicolumn{1}{c|}{$+0.10$} & 
\multicolumn{1}{c|}{0.01} & \multicolumn{1}{c|}{3282} & \multicolumn{1}{c|}{$%
-19.650$} & $0.210$ \\ \hline
\multicolumn{1}{|c|}{XVI$_{19}$} & \multicolumn{1}{c|}{107} & 
\multicolumn{1}{c|}{$0.6\leq z<1.4$} & \multicolumn{1}{c|}{$0.837$} & 
\multicolumn{1}{c|}{24.339} & \multicolumn{1}{c|}{$+0.15$} & 
\multicolumn{1}{c|}{0.02} & \multicolumn{1}{c|}{3658} & \multicolumn{1}{c|}{$%
-19.719$} & $0.202$ \\ \hline
\multicolumn{1}{|c|}{XVI$_{20}$} & \multicolumn{1}{c|}{75} & 
\multicolumn{1}{c|}{$0.7\leq z<1.4$} & \multicolumn{1}{c|}{$0.919$} & 
\multicolumn{1}{c|}{24.627} & \multicolumn{1}{c|}{$+0.28$} & 
\multicolumn{1}{c|}{0.03} & \multicolumn{1}{c|}{4075} & \multicolumn{1}{c|}{$%
-19.773$} & $0.188$ \\ \hline
\multicolumn{1}{|c|}{XVI$_{21}$} & \multicolumn{1}{c|}{58} & 
\multicolumn{1}{c|}{$0.8\leq z<1.4$} & \multicolumn{1}{c|}{$0.969$} & 
\multicolumn{1}{c|}{24.784} & \multicolumn{1}{c|}{$+0.26$} & 
\multicolumn{1}{c|}{0.04} & \multicolumn{1}{c|}{4323} & \multicolumn{1}{c|}{$%
-19.789$} & $0.193$ \\ \hline
\multicolumn{1}{|c|}{\textbf{XVI}$_1$} & \multicolumn{1}{c|}{\textbf{48}} & 
\multicolumn{1}{c|}{$0.83\leq z<1.4$} & \multicolumn{1}{c|}{$1.001$} & 
\multicolumn{1}{c|}{24.836} & \multicolumn{1}{c|}{$+0.29$} & 
\multicolumn{1}{c|}{0.03} & \multicolumn{1}{c|}{4409} & \multicolumn{1}{c|}{$%
-19.852$} & $0.175$ \\ \hline
\end{tabular}

\vspace{.2in}

By an unweighted fitting of the 13 normal points of Table 0, plotting the
listed $\langle \left| \Delta M\right| \rangle $ values versus the
corresponding $\langle D_z\rangle $ as shown in Appendix Figure 1, one can
draw an important relationship, according to the following two formulations: 
\begin{equation}
\langle \left| \Delta M\right| \rangle =0.40(\pm 0.01)-0.000053(\pm
0.000004)\cdot \langle D_z\rangle
\end{equation}
\begin{equation}
\langle \left| \Delta M\right| \rangle =1.42(\pm 0.05)-0.15(\pm 0.01)\cdot
\ln (\langle D_z\rangle )
\end{equation}
which are well confirmed by the corresponding two unweighted fittings of the
249 $\left| \Delta M\right| $ from Table 3abcdefghi, as follows: 
\begin{equation}
\langle \left| \Delta M\right| \rangle =0.40(\pm 0.04)-0.000052(\pm
0.000013)\cdot D_z
\end{equation}
\begin{equation}
\langle \left| \Delta M\right| \rangle =1.45(\pm 0.26)-0.15(\pm 0.04)\cdot
\ln (D_z)
\end{equation}

Both the previous correlations, (19) and (20), are shown in Appendix Figure
2, as fitting lines which run very near to each other at all the 249 SNe
Hubble depths. At the same time, plotting the $\langle M_z\rangle $ values
versus the corresponding $\langle D_z\rangle $ for the 13 normal points of
Table 0, as shown in Appendix Figure 3, allows a quick check of the fitting
curve II (2$^{nd}$ order), whose ECM\ equation 
\begin{equation}
\langle M_z\rangle \cong -\text{18.62E00}-\text{4E-04}\langle D_z\rangle +%
\text{3E-08}\langle D_z\rangle ^2
\end{equation}
results to agree with the paper IX eq. (22), that based on 30 (practically
model independent) normal points, including all the 398 SNe Ia with $z$ and $%
m_B^{\max }$ listed in Table 11 of the $SCPU$ compilation (Kowalski et al.
2008). Let us recall the paper IX values $d_0=-$18.77E00, $d_1=-$3.318E-04, $%
d_2=$1.957E-08, also used in paper XV and in the present dipole analysis,
according to eqs. (7)(8)(9). Other two fitting curves, III and IV (of 3$%
^{rd} $ and 4$^{th}$ order respectively), are represented in Appendix
Figures 4 and 5, where the relative equations show the peculiarity of a
systematic reduction in modulus of the zero order coefficient, according to
the corresponding values: $-$18.77, $-$18.62, $-$18.51,$-$18.36.

Appendix Figure 6 shows the plot and cubic fitting of the 249 SNe $M_z$
listed in Table 3abcdefghi against the corresponding $D_z$. It is important
to remark that here the zero order coefficient, as $-$18.15, has a value in
modulus smaller than the previous ones.

From the 249 $\Delta M$ values of Table 3abcdefghi, even a few rough
correlations of $\left| \Delta M\right| $ with $\cos \gamma $ seem to come
out; these are: 
\begin{equation}
\langle \left| \Delta M\right| \rangle \approx 0.26-0.05\cdot \cos \gamma %
\hspace{.3in}at\hspace{.3in}0.2<z<1.4\hspace{.6in}
\end{equation}
\begin{equation}
\langle \left| \Delta M\right| \rangle \approx 0.34-0.10\cdot \cos \gamma %
\hspace{.3in}at\hspace{.3in}0.2<z\leq 0.5\hspace{.3in}\text{(see Appendix
Figure 7)}
\end{equation}
\begin{equation}
\langle \left| \Delta M\right| \rangle \approx 0.21-0.00\cdot \cos \gamma %
\hspace{.3in}at\hspace{.3in}0.5<z<1.4\hspace{.6in}
\end{equation}

A further remark about the data of Table 0 regards the $7^{th}$ column,
where the small positive values of $\frac{\Delta D}D$ may indicate a
systematic scattering of $\langle z\rangle $ from $z_0$. Furthermore $\Delta
D$ has here the same behaviour as in Table 1b of the parallel paper XV. For
instance the sample XVI$_1$ of Table 0, whose $\langle D_z\rangle =cz_0/H_0$%
, gives $z_0=1.029$ and $\langle z\rangle -z_0=-0.028$ or $\left| \frac{%
\Delta z}z\right| \cong 0.03$. At the same time the dipole tests A1 and B1
of paper XV, with $\langle D\rangle =cz_0/H_0$, give $z_0=1.032$ and $%
z_0=1.045$, that is the corresponding $\langle z\rangle -z_0=-0.031$ and $%
\langle z\rangle -z_0=-0.044$, or $\left| \frac{\Delta z}z\right| \cong 0.03$
and $0.04$, respectively.

\subsection{ECM dipole tests weighted by $w_i\propto \left| \Delta M\right|
^{-i}$}

The results of the dipole test based on the weighted ECM Hubble law (16),
applied to each supernova of Table 3abcdefghi with weight $w_i\propto \left|
\Delta M\right| ^{-i}$ , are listed in Table 1. Here the $9$ columns present
three ECM dipole solutions for each SNe sample, with $i=0,1,2$ respectively,
as follows: Test identification name (TID); sample ordinal number; number N
of the sample SNe; the fitting standard deviation $s(w_0)$ in H.u. of the
unweighted ECM dipole test carried out on the line sample and the resulting
angular coefficient $a_0$ of eq. (16) with its standard deviation, in H.u.,
corresponding to the weight applied $w_0=1$ to each sample SNe; the standard
deviation $s(w_1)$ in H.u. of the fitting with $w_1=\left| \Delta M\right|
^{-1}$together with the resulting $a_0$ in H.u.; the standard deviation $%
s(w_2)$ in H.u. of the fitting with $w_2=\left| \Delta M\right| ^{-2}$%
together with the resulting $a_0$ in H.u..

\textbf{Table 1}

\vspace{.1in}

\begin{tabular}{|c|c|c||c|c||c|c||c|c|}
\hline
TID & Sample & N & $s(w_0)$ & $a_0$ & $s(w_1)$ & $a_0$ & $s(w_2)$ & $a_0$ \\ 
\hline
W11 & XVI$_{11}$ & 50 & 12.52 & $-1.2\pm 5.1$ & 4.40 & $11.3$ & 0.45 & $12.7$
\\ \hline
W12 & XVI$_{12}$ & 101 & 11.62 & $-2.9\pm 3.7$ & 4.63 & $11.0$ & 0.57 & $%
12.8 $ \\ \hline
W13 & XVI$_{13}$ & 142 & 10.95 & $-1.4\pm 3.0$ & 4.09 & $11.0$ & 0.57 & $%
12.7 $ \\ \hline
W14 & XVI$_{14}$ & 174 & 10.92 & $1.8\pm 2.8$ & 4.20 & $11.3$ & 0.62 & $12.7$
\\ \hline
W15 & XVI$_{15}$ & 192 & 10.62 & $2.3\pm 2.6$ & 4.18 & $11.3$ & 0.65 & $12.7$
\\ \hline
W16 & XVI$_{16}$ & 215 & 10.42 & $3.1\pm 2.4$ & 3.73 & $11.3$ & 0.45 & $12.7$
\\ \hline
W0 & XVI & 249 & 10.11 & $4.3\pm 2.2$ & 3.75 & $11.6$ & 0.48 & $12.75$ \\ 
\hline
W17 & XVI$_{17}$ & 200 & 9.409 & $6.6\pm 2.4$ & 3.60 & $11.9$ & 0.49 & $12.8$
\\ \hline
W18 & XVI$_{18}$ & 149 & 8.547 & $10.8\pm 2.6$ & 3.22 & $12.4$ & 0.44 & $%
12.7 $ \\ \hline
W19 & XVI$_{19}$ & 107 & 8.254 & $14.1\pm 3.0$ & 3.26 & $13.5$ & 0.40 & $%
13.3 $ \\ \hline
W20 & XVI$_{20}$ & 75 & 7.618 & $11.8\pm 3.3$ & 2.82 & $12.9$ & 0.33 & $13.3$
\\ \hline
W21 & XVI$_{21}$ & 58 & 7.794 & $12.7\pm 3.9$ & 2.65 & $13.5$ & 0.29 & $13.3$
\\ \hline
W1 & \textbf{XVI}$_1$ & \textbf{48} & 7.109 & $14.4\pm 3.9$ & 2.39 & $14.1$
& 0.26 & $13.4$ \\ \hline
\end{tabular}

\vspace{.2in}

At first sight the results in Table 1 seem to suggest that only the high
values of $\langle \left| \Delta M\right| \rangle $ in the 10$^{th}$ column
of Table 0, corresponding to $z\lesssim 0.5$, are significantly affecting
the unweighted ECM Hubble law (4$^{th\text{ }}$and 5$^{th\text{ }}$columns
of Table 1) with $D=D_{\langle M\rangle }$.\ On the other hand only the
weights $w_2=\left| \Delta M\right| ^{-2}$ give $a_0$ the exact ECM standard
value $12.7$ at $z\lesssim 0.5$; this means the ECM agrees with the adopted $%
\langle M\rangle $ $=d_0+d_1D_z+d_2D_z^2$ at that $z$ range. In other words
the solutions in Table 1 represent a further successful check of the
expansion center model at any Hubble depth of the supernovae Ia. As an
illustration, the dipole diagram of the unweighted test W18 is reported in
Appendix Figure 8. This ECM dipole test, referring to the SNe of Table
3abcdefghi with $z\geq 0.5$, is graphically represented by the fitted plot
of 149 values of $Y_{W18}=\frac{cz}{D_{\langle M\rangle }}-H_0$ against each
corresponding value of $-X_z$ (cf. section 4). Appendix Figure 9 represents
the same diagram of 3 normal points $\langle Y\rangle $ versus the
corresponding $\langle -X_z\rangle $, which include: $74$ SNe at the range $%
-X_z<0$; $52$ SNe at $0<$ $-X_z<0.25$ and $23$ SNe at $-X_z>0.25$.

\subsection{ECM dipole test based on $\Delta M\equiv 0$}

Within the previous dipole test, when one assumes $M\equiv \langle M\rangle $
or $\Delta M\equiv 0$, eq. (14) immediately leads to the identity $%
D_{\langle M\rangle }\equiv D_z$, that is a Hubble depth $D$ which should
agree with both the ECM Hubble law (1) and the Hubble Magnitude formulation
of eq. (10). This is the case of the $1^{st}$ type dipole test in paper XV,
according to the paper IX procedure, here integrated by the ECM formulae and
summarized as follows:

\begin{equation}
M-\langle M\rangle =\Delta M\rightarrow 0\Rightarrow M\equiv \langle
M\rangle =d_2D^2+d_1D+d_0=m-5\log D_C-25
\end{equation}
\begin{equation}
D_C=D\cdot (1+z)\Rightarrow d_2D^2+d_1D+d_0+5\log D=m-5\log (1+z)-25
\end{equation}
\begin{equation}
\left[ z,m,d_0,d_1,d_2\Rightarrow D\right]
\end{equation}
\begin{equation}
D=\frac{xc}{3H_0}\left( \frac{1+x}{1-x}\right) \Rightarrow x=x(D)\Rightarrow
X=X(x,\cos \gamma )=X(D,\cos \gamma )
\end{equation}
\begin{equation}
cz=(H_0-a_0X)\cdot D\Rightarrow Y=\left( \frac{cz}D-H_0\right) \Rightarrow
Y\rightarrow -X\Rightarrow a_0
\end{equation}

Eq. (29) has been checked again on the pilot sample XVI, using all the $cz$
and $D$ values listed in Table 3abcdefghi in the paper IX appendix. The $%
\cos \gamma $ introduction allows a further ECM dipole test on the same 13 $%
z $ bins in Table 0. The resulting angular coefficient $a_0$ of each
unweighted dipole test and the corresponding standard deviation $s$, in
H.u., are reported in the last two columns of Table 2; here the first column
is the TID names, as the continuation of the A series in paper XV. Also this
ECM dipole test based on $\Delta M\equiv 0$, as the results of Table 2 show
when compared with those of Table 1, gives evidence for the perturbative $%
\Delta M$ effect at $z\lesssim 0.5$. As in the previous section, Appendix
Figure 10 presents the dipole diagram of the test A18, as a plot of $Y_{A18}$
versus $-X$. Appendix Figure 11 represents the same diagram of 3 normal
points $\langle Y\rangle $ versus the corresponding $\langle -X\rangle $,
which include: $74$ SNe at the range $-X<0$; $52$ SNe at $0<$ $-X<0.25$ and $%
23$ SNe at $-X>0.25$.

\textbf{Table 2}

\vspace{.1in}

\begin{tabular}{|cccccc|}
\hline
\multicolumn{1}{|c|}{TID} & \multicolumn{1}{c|}{Sample} & 
\multicolumn{1}{c|}{N} & \multicolumn{1}{c|}{$z$ bin} & \multicolumn{1}{c|}{$%
s$} & $a_0$ \\ \hline
\multicolumn{1}{|c|}{A11} & \multicolumn{1}{c|}{XVI$_{11}$} & 
\multicolumn{1}{c|}{50} & \multicolumn{1}{c|}{$0.2<z\leq 0.4$} & 
\multicolumn{1}{c|}{14.89} & $-2\pm 6$ \\ \hline
\multicolumn{1}{|c|}{A12} & \multicolumn{1}{c|}{XVI$_{12}$} & 
\multicolumn{1}{c|}{101} & \multicolumn{1}{c|}{$0.2<z\leq 0.5$} & 
\multicolumn{1}{c|}{14.35} & $-5\pm 5$ \\ \hline
\multicolumn{1}{|c|}{A13} & \multicolumn{1}{c|}{XVI$_{13}$} & 
\multicolumn{1}{c|}{142} & \multicolumn{1}{c|}{$0.2<z\leq 0.6$} & 
\multicolumn{1}{c|}{13.78} & $-4\pm 4$ \\ \hline
\multicolumn{1}{|c|}{A14} & \multicolumn{1}{c|}{XVI$_{14}$} & 
\multicolumn{1}{c|}{174} & \multicolumn{1}{c|}{$0.2<z\leq 0.7$} & 
\multicolumn{1}{c|}{13.93} & $0\pm 4$ \\ \hline
\multicolumn{1}{|c|}{A15} & \multicolumn{1}{c|}{XVI$_{15}$} & 
\multicolumn{1}{c|}{192} & \multicolumn{1}{c|}{$0.2<z\leq 0.8$} & 
\multicolumn{1}{c|}{13.62} & $1\pm 4$ \\ \hline
\multicolumn{1}{|c|}{A16} & \multicolumn{1}{c|}{XVI$_{16}$} & 
\multicolumn{1}{c|}{215} & \multicolumn{1}{c|}{$0.2<z\leq 0.9$} & 
\multicolumn{1}{c|}{13.43} & $1\pm 4$ \\ \hline
\multicolumn{1}{|c|}{A0} & \multicolumn{1}{c|}{XVI} & \multicolumn{1}{c|}{249
} & \multicolumn{1}{c|}{$0.2<z<1.4$} & \multicolumn{1}{c|}{13.15} & $3\pm 3$
\\ \hline
\multicolumn{1}{|c|}{A17} & \multicolumn{1}{c|}{XVI$_{17}$} & 
\multicolumn{1}{c|}{200} & \multicolumn{1}{c|}{$0.4\leq z<1.4$} & 
\multicolumn{1}{c|}{12.73} & $4\pm 4$ \\ \hline
\multicolumn{1}{|c|}{A18} & \multicolumn{1}{c|}{XVI$_{18}$} & 
\multicolumn{1}{c|}{149} & \multicolumn{1}{c|}{$0.5\leq z<1.4$} & 
\multicolumn{1}{c|}{11.94} & $9.5\pm 3.6$ \\ \hline
\multicolumn{1}{|c|}{A19} & \multicolumn{1}{c|}{XVI$_{19}$} & 
\multicolumn{1}{c|}{107} & \multicolumn{1}{c|}{$0.6\leq z<1.4$} & 
\multicolumn{1}{c|}{11.72} & $13.8\pm 4.2$ \\ \hline
\multicolumn{1}{|c|}{A20} & \multicolumn{1}{c|}{XVI$_{20}$} & 
\multicolumn{1}{c|}{75} & \multicolumn{1}{c|}{$0.7\leq z<1.4$} & 
\multicolumn{1}{c|}{11.01} & $9.6\pm 4.8$ \\ \hline
\multicolumn{1}{|c|}{A21} & \multicolumn{1}{c|}{XVI$_{21}$} & 
\multicolumn{1}{c|}{58} & \multicolumn{1}{c|}{$0.8\leq z<1.4$} & 
\multicolumn{1}{c|}{11.24} & $11.1\pm 5.6$ \\ \hline
\multicolumn{1}{|c|}{A1} & \multicolumn{1}{c|}{\textbf{XVI}$_1$} & 
\multicolumn{1}{c|}{\textbf{48}} & \multicolumn{1}{c|}{$0.83\leq z<1.4$} & 
\multicolumn{1}{c|}{10.50} & $12.6\pm 5.8$ \\ \hline
\end{tabular}

\vspace{.2in}

\subsection{The SNe $\Delta M$ effect}

All the previous dipole tests seem to give $\Delta M$ a crucial and
macroscopic perturbation role, within the adopted expansion center model.
What might be the nature of such a $\Delta M$ ? Here, at least two origins
have to be taken into account, intrinsic or statistical. While the first has
to do with the physics and gravitation of the supernova itself, the latter
may be due both to selection effects and limits in the model, which is
formally correct when applied to the very nearby Universe with a rigid
rotation (cf. paper VII and section 7.4 of paper I). In fact the ECM dipoles
were well confirmed in the nearby Universe (cf. papers I-II and also Lorenzi
1991-93), without using supernovae; further confirmation came only from the
far Abell clusters, the 66 of Richness 3, at $z\lesssim 0.3$ and $\langle
z\rangle \cong 0.2$ (cf. paper V and also Lorenzi 1994). A first successful
dipole test on SNe Ia was carried out through two historic and accurate SCP
samples, by P99 and K03, at the average redshift $\langle z\rangle =0.5$
(cf. paper VI). The latest ECM confirmation refers to the Deep Universe, at $%
0.2<z\lesssim 1.4$, as shown in this work and in the parallel paper XV.
Consequently, the present disagreement of the unweighted SNe dipoles at $%
z\lesssim 0.5$ is very likely due to the perturbation effect of the SNe $%
\Delta M$, producing both an intrinsic and statistical interference.

\section{ECM values from 249 High-$z$ SCP Union SNe Ia}

This section is devoted to presenting 1743 data in Hubble units, calculated
for 249 High-$z$\ SCP Union supernovae, according to the expansion center
model. In particular the first three columns of Table 3abcdefghi list below
in order: Supernova name as reported in the 2008 SCP Union paper ($SCPU$:
Kowalski et al. 2008); redshift $z_{SCP}$ of supernova or host galaxy as
listed in $SCPU$, but rounded off to the third decimal place as the CMB
reference affects the value for about 0.001 on average (cf. paper IX);
supernova magnitude $m_{SCPU}$ as $m_B^{\max }$ value listed in $SCPU$,
without standard deviation. The fourth column holds the calculated value of $%
-\cos \gamma $, according to eq. (16) of paper XV, after introducing the SNe
R.A. $\alpha $ and Decl. $\delta $, those listed in paper XV Table 5abc or
in the $SCPU$\ reference papers (cf. Harvard-IAU, Riess et al. 2007, Astier
et al. 2006, Riess et al. 2004, Miknaitis et al. 2007). The following four
columns are all dedicated to as many ECM values, here called $r_z$,$D_z$,$%
M_z $,$-X_z$ in that directly coming from eq. (1) with the ECM standard
values $H_0=70$ H.u. and $a_0=12.7$ H.u. applied. Lastly, the $9^{th}$
column reports the integer value of the Hubble ratio $\frac{cz}{D_{\langle
M\rangle }}$, with $D_{\langle M\rangle }$ calculated through eq. (13),
while column $10$ lists the crucial value of $\Delta M=M_z-\langle M\rangle $%
, which represents the ECM scattering of the SN Hubble Magnitude with
respect the average value $\langle M\rangle $ of eq. (7).

\subsection{Evidence for intrinsic SNe $\Delta M$}

Concerning Table 3 below, one can attempt to search for the possible
intrinsic nature of some large $\Delta M$. First of all one's attention must
fall on those SNe which present very high $\Delta M$, for example $\Delta
M\geq +1.0$, as in the case of the following SNe: 03D4au with $\Delta
M=+1.03 $; g055 with $\Delta M=+1.39$; g142 with $\Delta M=+0.98$; k485 with 
$\Delta M=+1.26$. Another simple and more powerful procedure is based on the
comparison of only a few pairs of supernovae which, with almost the same
redshift and position on the celestial sphere, show very different $\Delta M$%
. To this end we also need to check the SNe coordinates, as the same ECM $%
\cos \gamma $ refers to the same hemisphere, not necessarily to the same
position in the sky. In Table 3 we can find a few of such SNe couples with
at least one $\Delta M>0.5$ and $\Delta (\Delta M)>0.5$ (as an example we
cite the couple 05Zwi-2002hr). The aforementioned evidence for a possible
intrinsic origin of many SNe $\Delta M$ is very important, in that it
appears to represent a crucial proof, in accordance with the expansion
center model, against the common assumption of using the supernovae SNe Ia
as good standard candles. In particular, at present it results that \textbf{%
the individual SNe Ia are not usable standard candles}.

\newpage
\ \textbf{Table 3a}

\vspace*{0.05in}

\begin{tabular}{|llllllll|l|l|}
\hline
\multicolumn{1}{|l|}{Name} & \multicolumn{1}{l|}{$z_{SCP}$} & 
\multicolumn{1}{l|}{$m_{SCPU}$} & \multicolumn{1}{l|}{$-\cos \gamma $} & 
\multicolumn{1}{l|}{$r_z$} & \multicolumn{1}{l|}{$D_z$} & 
\multicolumn{1}{l|}{$M_z$} & $-X_z$ & \multicolumn{1}{l|}{$\frac{cz}{%
D_{\langle M\rangle }}$} & $\Delta M$ \\ \hline
\multicolumn{1}{|l|}{1996h} & \multicolumn{1}{l|}{0.620} & 
\multicolumn{1}{l|}{23.50} & \multicolumn{1}{l|}{-0.5301} & 
\multicolumn{1}{l|}{794} & \multicolumn{1}{l|}{2784} & \multicolumn{1}{l|}{
-19.77} & -0.260 & \multicolumn{1}{|c|}{74} & -0.23 \\ \hline
\multicolumn{1}{|l|}{1996i} & \multicolumn{1}{l|}{0.570} & 
\multicolumn{1}{l|}{23.40} & \multicolumn{1}{l|}{-0.6082} & 
\multicolumn{1}{l|}{771} & \multicolumn{1}{l|}{2582} & \multicolumn{1}{l|}{
-19.64} & -0.305 & \multicolumn{1}{|c|}{71} & -0.14 \\ \hline
\multicolumn{1}{|l|}{1996j} & \multicolumn{1}{l|}{0.300} & 
\multicolumn{1}{l|}{22.03} & \multicolumn{1}{l|}{-0.6961} & 
\multicolumn{1}{l|}{583} & \multicolumn{1}{l|}{1388} & \multicolumn{1}{l|}{
-19.25} & -0.415 & \multicolumn{1}{|c|}{67} & -0.06 \\ \hline
\multicolumn{1}{|l|}{1996k} & \multicolumn{1}{l|}{0.380} & 
\multicolumn{1}{l|}{22.64} & \multicolumn{1}{l|}{-0.8527} & 
\multicolumn{1}{l|}{658} & \multicolumn{1}{l|}{1783} & \multicolumn{1}{l|}{
-19.32} & -0.475 & \multicolumn{1}{|c|}{64} & -0.02 \\ \hline
\multicolumn{1}{|l|}{1996u} & \multicolumn{1}{l|}{0.430} & 
\multicolumn{1}{l|}{22.61} & \multicolumn{1}{l|}{-0.7171} & 
\multicolumn{1}{l|}{690} & \multicolumn{1}{l|}{1981} & \multicolumn{1}{l|}{
-19.65} & -0.388 & \multicolumn{1}{|c|}{75} & -0.30 \\ \hline
\multicolumn{1}{|l|}{1995ao} & \multicolumn{1}{l|}{0.240} & 
\multicolumn{1}{l|}{21.60} & \multicolumn{1}{l|}{+0.0241} & 
\multicolumn{1}{l|}{496} & \multicolumn{1}{l|}{1024} & \multicolumn{1}{l|}{
-18.92} & +0.016 & \multicolumn{1}{|c|}{65} & +0.17 \\
\hline
\end{tabular}

\newpage\ \textbf{Table 3b}

\vspace*{0.05in}

\begin{tabular}{|llllllll|l|l|}
\hline
\multicolumn{1}{|l|}{Name} & \multicolumn{1}{l|}{$z_{SCP}$} & 
\multicolumn{1}{l|}{$m_{SCPU}$} & \multicolumn{1}{l|}{$-\cos \gamma $} & 
\multicolumn{1}{l|}{$r_z$} & \multicolumn{1}{l|}{$D_z$} & 
\multicolumn{1}{l|}{$M_z$} & $-X_z$ & \multicolumn{1}{l|}{$\frac{cz}{%
D_{\langle M\rangle }}$} & $\Delta M$ \\ \hline
\multicolumn{1}{|l|}{1996cf} & \multicolumn{1}{l|}{0.570} & 
\multicolumn{1}{l|}{23.31} & \multicolumn{1}{l|}{-0.7708} & 
\multicolumn{1}{l|}{776} & \multicolumn{1}{l|}{2624} & \multicolumn{1}{l|}{
-19.76} & -0.384 & \multicolumn{1}{|c|}{73} & -0.26 \\ \hline
\multicolumn{1}{|l|}{1995ba} & \multicolumn{1}{l|}{0.388} & 
\multicolumn{1}{l|}{22.55} & \multicolumn{1}{l|}{-0.9117} & 
\multicolumn{1}{l|}{666} & \multicolumn{1}{l|}{1831} & \multicolumn{1}{l|}{
-19.48} & -0.504 & \multicolumn{1}{|c|}{68} & -0.16 \\ \hline
\multicolumn{1}{|l|}{1995az} & \multicolumn{1}{l|}{0.450} & 
\multicolumn{1}{l|}{22.61} & \multicolumn{1}{l|}{-0.3184} & 
\multicolumn{1}{l|}{691} & \multicolumn{1}{l|}{1987} & \multicolumn{1}{l|}{
-19.69} & -0.172 & \multicolumn{1}{|c|}{79} & -0.34 \\ \hline
\multicolumn{1}{|l|}{1995ay} & \multicolumn{1}{l|}{0.480} & 
\multicolumn{1}{l|}{23.06} & \multicolumn{1}{l|}{-0.0072} & 
\multicolumn{1}{l|}{702} & \multicolumn{1}{l|}{2060} & \multicolumn{1}{l|}{
-19.36} & -0.004 & \multicolumn{1}{|c|}{70} & +0.01 \\ \hline
\multicolumn{1}{|l|}{1995ax} & \multicolumn{1}{l|}{0.615} & 
\multicolumn{1}{l|}{23.22} & \multicolumn{1}{l|}{+0.1194} & 
\multicolumn{1}{l|}{774} & \multicolumn{1}{l|}{2607} & \multicolumn{1}{l|}{
-19.90} & +0.060 & \multicolumn{1}{|c|}{85} & -0.40 \\ \hline
\multicolumn{1}{|l|}{1995aw} & \multicolumn{1}{l|}{0.400} & 
\multicolumn{1}{l|}{22.18} & \multicolumn{1}{l|}{+0.1244} & 
\multicolumn{1}{l|}{642} & \multicolumn{1}{l|}{1691} & \multicolumn{1}{l|}{
-19.69} & +0.070 & \multicolumn{1}{|c|}{86} & -0.42 \\
\hline
\end{tabular}

\newpage\ \textbf{Table 3c}

\vspace*{0.05in}

\begin{tabular}{|llllllll|l|l|}
\hline
\multicolumn{1}{|l|}{Name} & \multicolumn{1}{l|}{$z_{SCP}$} & 
\multicolumn{1}{l|}{$m_{SCPU}$} & \multicolumn{1}{l|}{$-\cos \gamma $} & 
\multicolumn{1}{l|}{$r_z$} & \multicolumn{1}{l|}{$D_z$} & 
\multicolumn{1}{l|}{$M_z$} & $-X_z$ & \multicolumn{1}{l|}{$\frac{cz}{%
D_{\langle M\rangle }}$} & $\Delta M$ \\ \hline
\multicolumn{1}{|l|}{2001iw} & \multicolumn{1}{l|}{0.340} & 
\multicolumn{1}{l|}{22.10} & \multicolumn{1}{l|}{-0.3503} & 
\multicolumn{1}{l|}{608} & \multicolumn{1}{l|}{1510} & \multicolumn{1}{l|}{
-19.43} & -0.204 & \multicolumn{1}{|c|}{74} & -0.20 \\ \hline
\multicolumn{1}{|l|}{2001iv} & \multicolumn{1}{l|}{0.396} & 
\multicolumn{1}{l|}{22.47} & \multicolumn{1}{l|}{-0.9022} & 
\multicolumn{1}{l|}{671} & \multicolumn{1}{l|}{1861} & \multicolumn{1}{l|}{
-19.60} & -0.497 & \multicolumn{1}{|c|}{73} & -0.28 \\ \hline
\multicolumn{1}{|l|}{2001hy} & \multicolumn{1}{l|}{0.812} & 
\multicolumn{1}{l|}{24.95} & \multicolumn{1}{l|}{-0.9686} & 
\multicolumn{1}{l|}{885} & \multicolumn{1}{l|}{3772} & \multicolumn{1}{l|}{
-19.22} & -0.433 & \multicolumn{1}{|c|}{51} & +0.52 \\ \hline
\multicolumn{1}{|l|}{2001hx} & \multicolumn{1}{l|}{0.799} & 
\multicolumn{1}{l|}{24.78} & \multicolumn{1}{l|}{-0.9695} & 
\multicolumn{1}{l|}{881} & \multicolumn{1}{l|}{3721} & \multicolumn{1}{l|}{
-19.35} & -0.435 & \multicolumn{1}{|c|}{54} & +0.39 \\ \hline
\multicolumn{1}{|l|}{2001hu} & \multicolumn{1}{l|}{0.882} & 
\multicolumn{1}{l|}{24.91} & \multicolumn{1}{l|}{-0.9007} & 
\multicolumn{1}{l|}{907} & \multicolumn{1}{l|}{4067} & \multicolumn{1}{l|}{
-19.51} & -0.393 & \multicolumn{1}{|c|}{57} & +0.29 \\ \hline
\multicolumn{1}{|l|}{2001hs} & \multicolumn{1}{l|}{0.833} & 
\multicolumn{1}{l|}{24.26} & \multicolumn{1}{l|}{-0.3503} & 
\multicolumn{1}{l|}{877} & \multicolumn{1}{l|}{3671} & \multicolumn{1}{l|}{
-19.88} & -0.158 & \multicolumn{1}{|c|}{73} & -0.16 \\
\hline
\end{tabular}

\newpage\ \textbf{Table 3d}

\vspace*{0.05in}

\begin{tabular}{|llllllll|l|l|}
\hline
\multicolumn{1}{|l|}{Name} & \multicolumn{1}{l|}{$z_{SCP}$} & 
\multicolumn{1}{l|}{$m_{SCPU}$} & \multicolumn{1}{l|}{$-\cos \gamma $} & 
\multicolumn{1}{l|}{$r_z$} & \multicolumn{1}{l|}{$D_z$} & 
\multicolumn{1}{l|}{$M_z$} & $-X_z$ & \multicolumn{1}{l|}{$\frac{cz}{%
D_{\langle M\rangle }}$} & $\Delta M$ \\ \hline
\multicolumn{1}{|l|}{05Lan} & \multicolumn{1}{l|}{1.230} & 
\multicolumn{1}{l|}{26.02} & \multicolumn{1}{l|}{-0.6783} & 
\multicolumn{1}{l|}{993} & \multicolumn{1}{l|}{5531} & \multicolumn{1}{l|}{
-19.44} & -0.269 & \multicolumn{1}{|c|}{51} & +0.57 \\ \hline
\multicolumn{1}{|l|}{05Red} & \multicolumn{1}{l|}{1.190} & 
\multicolumn{1}{l|}{25.76} & \multicolumn{1}{l|}{-0.6782} & 
\multicolumn{1}{l|}{985} & \multicolumn{1}{l|}{5369} & \multicolumn{1}{l|}{
-19.59} & -0.272 & \multicolumn{1}{|c|}{55} & +0.40 \\ \hline
\multicolumn{1}{|l|}{05Spo} & \multicolumn{1}{l|}{0.839} & 
\multicolumn{1}{l|}{24.20} & \multicolumn{1}{l|}{-0.6779} & 
\multicolumn{1}{l|}{887} & \multicolumn{1}{l|}{3798} & \multicolumn{1}{l|}{
-20.02} & -0.302 & \multicolumn{1}{|c|}{75} & -0.27 \\ \hline
\multicolumn{1}{|l|}{05Str} & \multicolumn{1}{l|}{1.010} & 
\multicolumn{1}{l|}{25.03} & \multicolumn{1}{l|}{-0.6793} & 
\multicolumn{1}{l|}{940} & \multicolumn{1}{l|}{4564} & \multicolumn{1}{l|}{
-19.78} & -0.286 & \multicolumn{1}{|c|}{64} & +0.09 \\ \hline
\multicolumn{1}{|l|}{05Zwi} & \multicolumn{1}{l|}{0.521} & 
\multicolumn{1}{l|}{23.07} & \multicolumn{1}{l|}{+0.1235} & 
\multicolumn{1}{l|}{723} & \multicolumn{1}{l|}{2207} & \multicolumn{1}{l|}{
-19.56} & +0.065 & \multicolumn{1}{|c|}{76} & -0.15 \\ \hline
\multicolumn{1}{|l|}{2002dc} & \multicolumn{1}{l|}{0.475} & 
\multicolumn{1}{l|}{23.09} & \multicolumn{1}{l|}{-0.6785} & 
\multicolumn{1}{l|}{719} & \multicolumn{1}{l|}{2178} & \multicolumn{1}{l|}{
-19.44} & -0.357 & \multicolumn{1}{|c|}{67} & -0.04 \\
\hline
\end{tabular}

\newpage\ \textbf{Table 3e}

\vspace*{0.05in}

\begin{tabular}{|llllllll|l|l|}
\hline
\multicolumn{1}{|l|}{Name} & \multicolumn{1}{l|}{$z_{SCP}$} & 
\multicolumn{1}{l|}{$m_{SCPU}$} & \multicolumn{1}{l|}{$-\cos \gamma $} & 
\multicolumn{1}{l|}{$r_z$} & \multicolumn{1}{l|}{$D_z$} & 
\multicolumn{1}{l|}{$M_z$} & $-X_z$ & \multicolumn{1}{l|}{$\frac{cz}{%
D_{\langle M\rangle }}$} & $\Delta M$ \\ \hline
\multicolumn{1}{|l|}{03D3af} & \multicolumn{1}{l|}{0.532} & 
\multicolumn{1}{l|}{23.49} & \multicolumn{1}{l|}{-0.4855} & 
\multicolumn{1}{l|}{747} & \multicolumn{1}{l|}{2387} & \multicolumn{1}{l|}{
-19.33} & -0.249 & \multicolumn{1}{|c|}{63} & +0.13 \\ \hline
\multicolumn{1}{|l|}{03D1fc} & \multicolumn{1}{l|}{0.331} & 
\multicolumn{1}{l|}{21.80} & \multicolumn{1}{l|}{+0.1648} & 
\multicolumn{1}{l|}{584} & \multicolumn{1}{l|}{1393} & \multicolumn{1}{l|}{
-19.54} & +0.098 & \multicolumn{1}{|c|}{84} & -0.35 \\ \hline
\multicolumn{1}{|l|}{03D1bp} & \multicolumn{1}{l|}{0.346} & 
\multicolumn{1}{l|}{22.45} & \multicolumn{1}{l|}{+0.1674} & 
\multicolumn{1}{l|}{597} & \multicolumn{1}{l|}{1455} & \multicolumn{1}{l|}{
-19.01} & +0.099 & \multicolumn{1}{|c|}{65} & +0.20 \\ \hline
\multicolumn{1}{|l|}{04D4dw} & \multicolumn{1}{l|}{0.961} & 
\multicolumn{1}{l|}{24.57} & \multicolumn{1}{l|}{+0.9317} & 
\multicolumn{1}{l|}{889} & \multicolumn{1}{l|}{3824} & \multicolumn{1}{l|}{
-19.80} & +0.415 & \multicolumn{1}{|c|}{77} & -0.05 \\ \hline
\multicolumn{1}{|l|}{04D4an} & \multicolumn{1}{l|}{0.613} & 
\multicolumn{1}{l|}{24.02} & \multicolumn{1}{l|}{+0.9321} & 
\multicolumn{1}{l|}{751} & \multicolumn{1}{l|}{2418} & \multicolumn{1}{l|}{
-18.94} & +0.476 & \multicolumn{1}{|c|}{60} & +0.52 \\ \hline
\multicolumn{1}{|l|}{04D3nh} & \multicolumn{1}{l|}{0.340} & 
\multicolumn{1}{l|}{22.14} & \multicolumn{1}{l|}{-0.4821} & 
\multicolumn{1}{l|}{613} & \multicolumn{1}{l|}{1536} & \multicolumn{1}{l|}{
-19.43} & -0.280 & \multicolumn{1}{|c|}{73} & -0.19 \\
\hline
\end{tabular}

\newpage\ \textbf{Table 3f}

\vspace*{0.05in}

\begin{tabular}{|llllllll|l|l|}
\hline
\multicolumn{1}{|l|}{Name} & \multicolumn{1}{l|}{$z_{SCP}$} & 
\multicolumn{1}{l|}{$m_{SCPU}$} & \multicolumn{1}{l|}{$-\cos \gamma $} & 
\multicolumn{1}{l|}{$r_z$} & \multicolumn{1}{l|}{$D_z$} & 
\multicolumn{1}{l|}{$M_z$} & $-X_z$ & \multicolumn{1}{l|}{$\frac{cz}{%
D_{\langle M\rangle }}$} & $\Delta M$ \\ \hline
\multicolumn{1}{|l|}{04D1ak} & \multicolumn{1}{l|}{0.526} & 
\multicolumn{1}{l|}{23.63} & \multicolumn{1}{l|}{+0.1596} & 
\multicolumn{1}{l|}{725} & \multicolumn{1}{l|}{2221} & \multicolumn{1}{l|}{
-19.02} & +0.084 & \multicolumn{1}{|c|}{59} & +0.39 \\ \hline
\multicolumn{1}{|l|}{03D4gg} & \multicolumn{1}{l|}{0.592} & 
\multicolumn{1}{l|}{23.40} & \multicolumn{1}{l|}{+0.9331} & 
\multicolumn{1}{l|}{740} & \multicolumn{1}{l|}{2333} & \multicolumn{1}{l|}{
-19.45} & +0.482 & \multicolumn{1}{|c|}{76} & -0.01 \\ \hline
\multicolumn{1}{|l|}{03D4di} & \multicolumn{1}{l|}{0.905} & 
\multicolumn{1}{l|}{24.29} & \multicolumn{1}{l|}{+0.9334} & 
\multicolumn{1}{l|}{871} & \multicolumn{1}{l|}{3597} & \multicolumn{1}{l|}{
-19.89} & +0.423 & \multicolumn{1}{|c|}{82} & -0.18 \\ \hline
\multicolumn{1}{|l|}{03D4cx} & \multicolumn{1}{l|}{0.949} & 
\multicolumn{1}{l|}{24.50} & \multicolumn{1}{l|}{+0.9333} & 
\multicolumn{1}{l|}{885} & \multicolumn{1}{l|}{3772} & \multicolumn{1}{l|}{
-19.83} & +0.417 & \multicolumn{1}{|c|}{79} & -0.09 \\ \hline
\multicolumn{1}{|l|}{03D3cd} & \multicolumn{1}{l|}{0.461} & 
\multicolumn{1}{l|}{22.56} & \multicolumn{1}{l|}{-0.4916} & 
\multicolumn{1}{l|}{704} & \multicolumn{1}{l|}{2074} & \multicolumn{1}{l|}{
-19.85} & -0.263 & \multicolumn{1}{|c|}{83} & -0.47 \\ \hline
\multicolumn{1}{|l|}{03D3ay} & \multicolumn{1}{l|}{0.371} & 
\multicolumn{1}{l|}{22.20} & \multicolumn{1}{l|}{-0.4928} & 
\multicolumn{1}{l|}{639} & \multicolumn{1}{l|}{1675} & \multicolumn{1}{l|}{
-19.60} & -0.279 & \multicolumn{1}{|c|}{77} & -0.33 \\
\hline
\end{tabular}

\newpage\ \textbf{Table 3g}

\vspace*{0.05in}

\begin{tabular}{|llllllll|l|l|}
\hline
\multicolumn{1}{|l|}{Name} & \multicolumn{1}{l|}{$z_{SCP}$} & 
\multicolumn{1}{l|}{$m_{SCPU}$} & \multicolumn{1}{l|}{$-\cos \gamma $} & 
\multicolumn{1}{l|}{$r_z$} & \multicolumn{1}{l|}{$D_z$} & 
\multicolumn{1}{l|}{$M_z$} & $-X_z$ & \multicolumn{1}{l|}{$\frac{cz}{%
D_{\langle M\rangle }}$} & $\Delta M$ \\ \hline
\multicolumn{1}{|l|}{d058} & \multicolumn{1}{l|}{0.583} & 
\multicolumn{1}{l|}{23.59} & \multicolumn{1}{l|}{+0.4062} & 
\multicolumn{1}{l|}{749} & \multicolumn{1}{l|}{2402} & \multicolumn{1}{l|}{
-19.31} & +0.208 & \multicolumn{1}{|c|}{68} & +0.14 \\ \hline
\multicolumn{1}{|l|}{d084} & \multicolumn{1}{l|}{0.519} & 
\multicolumn{1}{l|}{23.64} & \multicolumn{1}{l|}{+0.1955} & 
\multicolumn{1}{l|}{720} & \multicolumn{1}{l|}{2185} & \multicolumn{1}{l|}{
-18.97} & +0.103 & \multicolumn{1}{|c|}{58} & +0.44 \\ \hline
\multicolumn{1}{|l|}{d085} & \multicolumn{1}{l|}{0.401} & 
\multicolumn{1}{l|}{22.48} & \multicolumn{1}{l|}{+0.7611} & 
\multicolumn{1}{l|}{624} & \multicolumn{1}{l|}{1593} & \multicolumn{1}{l|}{
-19.26} & +0.437 & \multicolumn{1}{|c|}{76} & -0.01 \\ \hline
\multicolumn{1}{|l|}{d087} & \multicolumn{1}{l|}{0.340} & 
\multicolumn{1}{l|}{21.91} & \multicolumn{1}{l|}{+0.3942} & 
\multicolumn{1}{l|}{585} & \multicolumn{1}{l|}{1397} & \multicolumn{1}{l|}{
-19.45} & +0.235 & \multicolumn{1}{|c|}{82} & -0.26 \\ \hline
\multicolumn{1}{|l|}{d089} & \multicolumn{1}{l|}{0.436} & 
\multicolumn{1}{l|}{22.50} & \multicolumn{1}{l|}{+0.1884} & 
\multicolumn{1}{l|}{666} & \multicolumn{1}{l|}{1831} & \multicolumn{1}{l|}{
-19.60} & +0.104 & \multicolumn{1}{|c|}{81} & -0.29 \\ \hline
\multicolumn{1}{|l|}{d093} & \multicolumn{1}{l|}{0.363} & 
\multicolumn{1}{l|}{21.89} & \multicolumn{1}{l|}{+0.1865} & 
\multicolumn{1}{l|}{611} & \multicolumn{1}{l|}{1525} & \multicolumn{1}{l|}{
-19.70} & +0.108 & \multicolumn{1}{|c|}{89} & -0.47 \\
\hline
\end{tabular}

\newpage
\textbf{Table 3h}

\vspace*{0.05in}

\begin{tabular}{|llllllll|l|l|}
\hline
\multicolumn{1}{|l|}{Name} & \multicolumn{1}{l|}{$z_{SCP}$} & 
\multicolumn{1}{l|}{$m_{SCPU}$} & \multicolumn{1}{l|}{$-\cos \gamma $} & 
\multicolumn{1}{l|}{$r_z$} & \multicolumn{1}{l|}{$D_z$} & 
\multicolumn{1}{l|}{$M_z$} & $-X_z$ & \multicolumn{1}{l|}{$\frac{cz}{%
D_{\langle M\rangle }}$} & $\Delta M$ \\ \hline
\multicolumn{1}{|l|}{f231} & \multicolumn{1}{l|}{0.619} & 
\multicolumn{1}{l|}{23.45} & \multicolumn{1}{l|}{+0.7531} & 
\multicolumn{1}{l|}{759} & \multicolumn{1}{l|}{2482} & \multicolumn{1}{l|}{
-19.57} & +0.382 & \multicolumn{1}{|c|}{78} & -0.10 \\ \hline
\multicolumn{1}{|l|}{f235} & \multicolumn{1}{l|}{0.422} & 
\multicolumn{1}{l|}{22.45} & \multicolumn{1}{l|}{+0.3897} & 
\multicolumn{1}{l|}{650} & \multicolumn{1}{l|}{1737} & \multicolumn{1}{l|}{
-19.51} & +0.219 & \multicolumn{1}{|c|}{81} & -0.23 \\ \hline
\multicolumn{1}{|l|}{f244} & \multicolumn{1}{l|}{0.540} & 
\multicolumn{1}{l|}{23.30} & \multicolumn{1}{l|}{+0.1861} & 
\multicolumn{1}{l|}{732} & \multicolumn{1}{l|}{2273} & \multicolumn{1}{l|}{
-19.42} & +0.097 & \multicolumn{1}{|c|}{71} & +.003 \\ \hline
\multicolumn{1}{|l|}{f308} & \multicolumn{1}{l|}{0.401} & 
\multicolumn{1}{l|}{23.07} & \multicolumn{1}{l|}{+0.1891} & 
\multicolumn{1}{l|}{641} & \multicolumn{1}{l|}{1686} & \multicolumn{1}{l|}{
-18.80} & +0.107 & \multicolumn{1}{|c|}{57} & +0.48 \\ \hline
\multicolumn{1}{|l|}{g005} & \multicolumn{1}{l|}{0.218} & 
\multicolumn{1}{l|}{21.32} & \multicolumn{1}{l|}{+0.7577} & 
\multicolumn{1}{l|}{447} & \multicolumn{1}{l|}{855} & \multicolumn{1}{l|}{
-18.77} & +0.509 & \multicolumn{1}{|c|}{67} & +0.27 \\ \hline
\multicolumn{1}{|l|}{g050} & \multicolumn{1}{l|}{0.633} & 
\multicolumn{1}{l|}{23.18} & \multicolumn{1}{l|}{+0.7625} & 
\multicolumn{1}{l|}{765} & \multicolumn{1}{l|}{2531} & \multicolumn{1}{l|}{
-19.90} & +0.384 & \multicolumn{1}{|c|}{91} & -0.42 \\
\hline
\end{tabular}

\newpage\textbf{Table 3i}

\vspace*{0.05in}

\begin{tabular}{|llllllll|l|l|}
\hline
\multicolumn{1}{|l|}{Name} & \multicolumn{1}{l|}{$z_{SCP}$} & 
\multicolumn{1}{l|}{$m_{SCPU}$} & \multicolumn{1}{l|}{$-\cos \gamma $} & 
\multicolumn{1}{l|}{$r_z$} & \multicolumn{1}{l|}{$D_z$} & 
\multicolumn{1}{l|}{$M_z$} & $-X_z$ & \multicolumn{1}{l|}{$\frac{cz}{%
D_{\langle M\rangle }}$} & $\Delta M$ \\ \hline
\multicolumn{1}{|l|}{k411} & \multicolumn{1}{l|}{0.564} & 
\multicolumn{1}{l|}{22.89} & \multicolumn{1}{l|}{+0.7643} & 
\multicolumn{1}{l|}{729} & \multicolumn{1}{l|}{2250} & \multicolumn{1}{l|}{
-19.84} & +0.399 & \multicolumn{1}{|c|}{91} & -0.42 \\ \hline
\multicolumn{1}{|l|}{k425} & \multicolumn{1}{l|}{0.274} & 
\multicolumn{1}{l|}{21.94} & \multicolumn{1}{l|}{+0.3917} & 
\multicolumn{1}{l|}{522} & \multicolumn{1}{l|}{1124} & \multicolumn{1}{l|}{
-18.84} & +0.246 & \multicolumn{1}{|c|}{64} & +0.28 \\ \hline
\multicolumn{1}{|l|}{k430} & \multicolumn{1}{l|}{0.582} & 
\multicolumn{1}{l|}{23.81} & \multicolumn{1}{l|}{+0.3826} & 
\multicolumn{1}{l|}{750} & \multicolumn{1}{l|}{2410} & \multicolumn{1}{l|}{
-19.10} & +0.196 & \multicolumn{1}{|c|}{61} & +0.36 \\ \hline
\multicolumn{1}{|l|}{k441} & \multicolumn{1}{l|}{0.680} & 
\multicolumn{1}{l|}{23.73} & \multicolumn{1}{l|}{+0.2204} & 
\multicolumn{1}{l|}{802} & \multicolumn{1}{l|}{2858} & \multicolumn{1}{l|}{
-19.68} & +0.107 & \multicolumn{1}{|c|}{75} & -0.12 \\ \hline
\multicolumn{1}{|l|}{k448} & \multicolumn{1}{l|}{0.401} & 
\multicolumn{1}{l|}{23.34} & \multicolumn{1}{l|}{+0.4038} & 
\multicolumn{1}{l|}{634} & \multicolumn{1}{l|}{1647} & \multicolumn{1}{l|}{
-18.48} & +0.230 & \multicolumn{1}{|c|}{51} & +0.79 \\ \hline
\multicolumn{1}{|l|}{k485} & \multicolumn{1}{l|}{0.416} & 
\multicolumn{1}{l|}{23.93} & \multicolumn{1}{l|}{+0.2254} & 
\multicolumn{1}{l|}{651} & \multicolumn{1}{l|}{1742} & \multicolumn{1}{l|}{
-18.03} & +0.126 & \multicolumn{1}{|c|}{40} & +1.26 \\
\hline
\end{tabular}

\newpage 

\section{A new ECM dipole test}

\textbf{\ }All the ECM papers, V, VI, IX, XV, and the previous sections of
paper XVI, have been developed by assuming the identity $D_C=D(1+z)\equiv
D_L $. As the Hubble depth $D$ represents an apparent distance at the
present epoch $t_0$ (cf. paper XV), so the related $D_L$ should be a
fictitious luminous distance $D_{FL}$, according to the arguments in paper
V. Consequently the resulting successful $M$, called Hubble Magnitude in
paper X and XV, may be different from the true absolute magnitude, though
the formula 
\begin{equation}
M=m-5\log D_L-25
\end{equation}
might make the necessary adjustement between the apparent magnitude $m$ and
the apparent distance $D_L$ in order to produce the correct value of the
absolute magnitude $M$. This is the problem: what luminosity distance $D_L$
is able to produce, not just a useful but likely fictitious value of $M$,
but the true $M$ ? The present section proposes the ECM \textbf{exploration
of the following }$D_L^{*}$\textbf{\ formula}: 
\begin{equation}
D_L^{*}=r\cdot (1+z)^2
\end{equation}

The previous $D_L^{*}$ equation differs from relativistic cosmology in that,
here, the light-space $r=-c\Delta t$ is a physical distance, representing
the space run by light during the past travel time $\Delta t=t-t_0$, in
place of the relativistic proper distance $r_{pr}$ at the emission epoch $t$
(cf. section 2 of papers VIII, IX, XV). However $r$ in light-time also
represents a measure of the past epoch $t$; in other words $r$ may be
considered to all intents and purposes the light-space distance of the
source at time $t$. That $r$, as $r_{z\text{ }}$, is the light space fitting
the ECM Hubble law (1). In this case the proposed experimental formulation
of the luminosity distance is eq. (31), here explored and tested on 249
High-z SCP Union supernovae, both to check the behaviour of the SNe Ia
absolute magnitude according to the expansion center model, where now the
High-z SNe Ia show low and slowly increasing average absolute magnitudes $%
\langle M^{*}\rangle $, and to \textbf{reconfirm the expansion dipole} of
the ECM Universe - with $\langle M^{*}\rangle \cong -18.01$ from all 249 SNe
- $\langle M^{*}\rangle \cong -18.02$ from 200 SNe at $z\geq 0.4$ - $\langle
M^{*}\rangle \cong -18.03$ from 149 SNe at $z\geq 0.5$ -.

The ECM test of eq. (31) is based on the 13 ECM normal points of the
corresponding samples in Table 0, here partially reproduced in Table 4, with
two new columns, $\langle r_z\rangle $ and the $\langle M^{*}\rangle $
resulting from the data of Table 3abcdefghi, as the sequence (32) reported
below shows: 
\begin{equation}
\mathbf{ECM}:\hspace{.2in}r=r_z\Rightarrow D_L^{*}\equiv
r_z(1+z)^2\Rightarrow \langle M^{*}\rangle =\langle m_B^{\max }\rangle
-5\langle \log D_L^{*}\rangle -25
\end{equation}

The linear fitting of the 13 normal $\langle M^{*}\rangle $ values of Table
4 versus the corresponding $\langle z\rangle $ listed in column 4$^{th}$%
gives the resulting relationship (33): 
\begin{equation}
\langle M^{*}\rangle =-17.83(\pm 0.01)-0.15(\pm 0.02)\cdot \langle z\rangle
\end{equation}

Hence a 2$^{nd}$ type dipole test (cf. section 3.2 of paper XV) based on eq.
(31) has been carried out through the following sequential steps: 
\begin{equation}
M^{*}=m-5\log \left[ r\cdot (1+z)^2\right] -25\Rightarrow
r=10^{0.2(m-M^{*})-5}/(1+z)^2
\end{equation}
\begin{equation}
\mathbf{ECM}:\hspace{.2in}x=\frac{3H_0r}c\Rightarrow D=r\cdot \left( \frac{%
1+x}{1-x}\right) \Rightarrow Y=\frac{cz}D-H_0=-\cos \gamma \cdot a^{*}(x)
\end{equation}
\begin{equation}
\left[ \gamma ,z,m,M^{*}(s_{Min})\right] \Rightarrow r=\frac{10^{0.2\left[
m-M^{*}(s_{Min})\right] -5}}{(1+z)^2}\Rightarrow Y\rightarrow -\cos \gamma
\Rightarrow a^{*}
\end{equation}
\begin{equation}
a^{*}(x)=a_0\cdot (1-x)^{\frac 13}/(1+x)\Rightarrow X=\cos \gamma \cdot
(1-x)^{\frac 13}/(1+x)\Rightarrow Y\rightarrow -X\Rightarrow a_0
\end{equation}

The least square procedure has been applied to all the 13 samples of Table
4. The obtained solutions of the unweighted fittings (36) and (37) are
listed in Table 5, for each new double dipole test \textbf{R} whose TID
number refers to the sample index of Table 4 column 1. The corresponding
rows present the resulting angular coefficients $a^{*}(x)$ and $a_0$,
preceded by the minimum value of the fitting standard deviations $s_{Min}$
and the related values $M^{*}(s_{Min})$, following the same order as in
Table 3 of paper V. In particular the expected value $a_{ECM}^{*}$ in column
2 derives from the $a^{*}(x)$ formula of (37) with $x=3H_0\langle r\rangle
/c $, $H_0=70$ and $a_0=12.7$ H.u. (cf. section 2.1), being $\langle
r\rangle $ the computed mean light distance of the sample according to (36).
The results of this new dipole test are important, though the \textbf{%
standard deviations }$s_{Min}$\textbf{\ have here more than doubled}. The
test gives a further ECM confirmation. Three large sets of High-z SNe Ia of
Table 5, the samples called \textbf{XVI -XVI}$_{17}$\textbf{-XVI}$_{18}$,
produce angular coefficients in accordance with those expected. Moreover the
mathematical means of all the 13 $a^{*}$ and $a_0$ values listed in Table 5
become $\langle a^{*}(x)\rangle =+6.1\pm 2.2$ and $\langle a_0\rangle =$ $%
+11.6\pm 3.8$, respectively, being $\langle a_{ECM}^{*}\rangle =+6.1\pm 0.2$
H.u.. Thus the present test, when compared with the previous ones based on $%
D_L=D\cdot (1+z)$ (cf. also the papers V-VI-IX-XV), clearly shows the ECM
dipole check as being independent from the inferred value of $M$, within the
limits of consistent formulations of the luminosity distance $D_L$. As in
the section 3.2 and 3.3, Appendix Figure 12 presents the dipole diagram of
the test \textbf{R18}, as a plot of $Y_{R18}$ versus $-X$. Appendix Figure
13 represents the same diagram of 3 normal points $\langle Y\rangle $ versus
the corresponding $\langle -X\rangle $, which include : $74$ SNe at the
range $-X<0$ ; $52$ SNe at $0<$ $-X<0.25$ and $23$ SNe at $-X>0.25$.

\newpage\ 

\textbf{Table 4}

\vspace{.06in}

\begin{tabular}{|c|c|c|c|c|c|c|c|}
\hline
Sample & N & $z$ bin & $\langle z\rangle $ & $\langle m\rangle $ & $\langle
D_z\rangle $ & $\langle r_z\rangle $ & $\langle M^{*}\rangle $ \\ \hline
XVI$_{11}$ & 50 & $0.2<z\leq 0.4$ & $0.322$ & 22.123 & 1425 & 577.6 & $%
-17.883$ \\ \hline
XVI$_{12}$ & 101 & $0.2<z\leq 0.5$ & $0.387$ & 22.534 & 1716 & 632.1 & $%
-17.869$ \\ \hline
XVI$_{13}$ & 142 & $0.2<z\leq 0.6$ & $0.434$ & 22.762 & 1893 & 663.5 & $%
-17.885$ \\ \hline
XVI$_{14}$ & 174 & $0.2<z\leq 0.7$ & $0.472$ & 22.928 & 2037 & 686.3 & $%
-17.902$ \\ \hline
XVI$_{15}$ & 192 & $0.2<z\leq 0.8$ & $0.499$ & 23.040 & 2152 & 701.3 & $%
-17.907$ \\ \hline
XVI$_{16}$ & 215 & $0.2<z\leq 0.9$ & $0.535$ & 23.195 & 2332 & 720.1 & $%
-17.902$ \\ \hline
\textbf{XVI} & \textbf{249} & $0.2<z<1.4$ & $0.607$ & 23.440 & 2651 & 750.4
& $-17.918$ \\ \hline
\textbf{XVI}$_{17}$ & \textbf{200} & $0.4\leq z<1.4$ & $0.677$ & 23.763 & 
2951 & 793.0 & $-17.928$ \\ \hline
\textbf{XVI}$_{18}$ & \textbf{149} & $0.5\leq z<1.4$ & $0.756$ & 24.052 & 
3282 & 830.4 & $-17.951$ \\ \hline
XVI$_{19}$ & 107 & $0.6\leq z<1.4$ & $0.837$ & 24.339 & 3658 & 865.6 & $%
-17.961$ \\ \hline
XVI$_{20}$ & 75 & $0.7\leq z<1.4$ & $0.919$ & 24.627 & 4075 & 898.9 & $%
-17.954$ \\ \hline
XVI$_{21}$ & 58 & $0.8\leq z<1.4$ & $0.969$ & 24.784 & 4323 & 914.4 & $%
-17.948$ \\ \hline
XVI$_1$ & 48 & $0.83\leq z<1.4$ & $1.001$ & 24.836 & 4409 & 924.9 & $-17.993$
\\ \hline
\end{tabular}

\vspace{.2in}

\textbf{Table 5}

\vspace{.06in}

\begin{tabular}{|c||c|c|c|c||c|c|c|}
\hline
TID & $a_{ECM}^{*}$ & $s_{Min}$ & $M^{*}(s_{Min})$ & $a^{*}(x)$ & $s_{Min}$
& $M^{*}(s_{Min})$ & $a_0$ \\ \hline
R11 & $+7.3$ & $23.190$ & $-18.02$ & $-2\pm 6$ & $23.209$ & $-18.01$ & $%
-2\pm 10$ \\ \hline
R12 & $+6.9$ & $23.157$ & $-17.98$ & $-4\pm 5$ & $23.202$ & $-17.98$ & $%
-5\pm 8$ \\ \hline
R13 & $+6.8$ & $22.313$ & $-17.99$ & $-3\pm 4$ & $22.346$ & $-17.99$ & $%
-4\pm 7$ \\ \hline
R14 & $+6.6$ & $22.662$ & $-18.01$ & $+1\pm 3$ & $22.654$ & $-18.01$ & $%
+2\pm 6$ \\ \hline
R15 & $+6.5$ & $22.372$ & $-18.01$ & $+1.7\pm 2.9$ & $22.368$ & $-18.01$ & $%
+3.7\pm 5.6$ \\ \hline
R16 & $+6.4$ & $22.551$ & $-17.99$ & $+2.8\pm 2.8$ & $22.562$ & $-17.99$ & $%
+4.9\pm 5.4$ \\ \hline
\textbf{R0} & $+6.2$ & $22.908$ & $-18.01$ & $+4.6\pm 2.6$ & $22.964$ & $%
-18.01$ & $+7.3\pm 5.2$ \\ \hline
\textbf{R17} & $+6.0$ & $22.794$ & $-18.02$ & $+6.6\pm 2.9$ & $22.887$ & $%
-18.01$ & $+11.1\pm 6.1$ \\ \hline
\textbf{R18} & $+5.8$ & $22.134$ & $-18.03$ & $+10.0\pm 3.1$ & $22.345$ & $%
-18.02$ & $+18.2\pm 6.9$ \\ \hline
R19 & $+5.5$ & $22.440$ & $-18.04$ & $+14\pm 4$ & $22.823$ & $-18.04$ & $%
+29\pm 9$ \\ \hline
R20 & $+5.3$ & $22.697$ & $-18.05$ & $+15\pm 5$ & $23.298$ & $-18.03$ & $%
+25\pm 11$ \\ \hline
R21 & $+5.2$ & $24.072$ & $-18.04$ & $+14\pm 6$ & $24.644$ & $-18.03$ & $%
+25\pm 13$ \\ \hline
R1 & $+5.2$ & $21.701$ & $-18.08$ & $+19\pm 6$ & $22.746$ & $-18.07$ & $%
+35\pm 14$ \\ \hline
\end{tabular}

\newpage

\section{Absolute magnitude analysis of the SCP Union supernovae}

After the preliminary magnitude analysis on the SCP Union data set in paper
IX, here a further more precise analysis is carried out so as to distinguish
the normal luminosity behaviour of the supernovae Ia of the deep Universe
from the SNe magnitude trend of the nearby Universe.

\subsection{Fitting 14 High-$z$ SNe $M$ normal points}

In the above sections we found evidence for a clear perturbation effect of
the SNe $\Delta M$ at $z\lesssim 0.5$. In order to avoid possible
interference effects, here a new model independent analysis of the normal
points in paper IX Table 2 is undertaken and limited to $14$ high-$z$ mean
Hubble Magnitudes $\langle M\rangle $, those with $z$-bin normal redshifts $%
\langle z\rangle >0.55$. If a first, second and third degree polynomial is
applied to the fitting of the $\langle M\rangle $ plot versus $\langle
z\rangle $, the statistical coefficients of determination $\mathbf{R}^{%
\mathbf{2}}$ are $0.9720,0.9967,0.9974,$ respectively. The best fitting is
clearly the cubic one. Therefore, after adopting the identity between the $z$%
-bin normal redshift and the central redshift $z_0$, that is 
\begin{equation}
\langle z\rangle \equiv z_0
\end{equation}
and the \textbf{normal equation of the Hubble Magnitude} 
\begin{equation}
\langle M\rangle =\langle m_B^{\max }\rangle -5\langle \log \left[
cz(1+z)\right] \rangle +5\log H_0-25
\end{equation}
the \textbf{line equation of the normal Hubble Magnitude }$\langle M\rangle $%
\textbf{\ as a function of the central redshift} $z_0$ becomes 
\begin{equation}
\langle M\rangle =A_0+A_1z_0+A_2z_0^2+A_3z_0^3
\end{equation}
with 
\begin{equation}
A_0=-17.96\hspace{.2in}A_1=-4.117\hspace{.2in}A_2=+3.197\hspace{.2in}%
A_3=-0.9463
\end{equation}
from the automatic cubic fitting (cf. Appendix Figure 14) whose coefficient
of determination $\mathbf{R}^{\mathbf{2}}=0.9974$.

Note that the previous eq. (2) of the normal points $\langle M\rangle $ is
the same normal $M$ equation (21) of paper IX, while the Hubble Magnitude $M$
of an individual source with redshift $z$ and apparent magnitude $m$ is by
definition 
\begin{equation}
M=m-5\log \left[ D\cdot (1+z)\right] -25
\end{equation}
where $D=cz/H_X=cz_0/H_0$ is the Hubble depth according to the expansion
center Universe (cf. the ECM papers V-VI-IX-XV).

Together with the successful cubic fitting (3) of $14$ high-$z$ normal
Hubble Magnitudes $\langle M\rangle $ versus the normal redshift $\langle
z\rangle $, it is possible to carry out a successful linear fitting of the
same $14$ $\langle M\rangle $ points versus the corresponding central light
space values $r=r(z_0)$ listed in column $8$ of paper IX Table 2. In this
case the normal Hubble Magnitude $\langle M\rangle $ is represented by the
equation 
\begin{equation}
\langle M\rangle =C_0+C_1r
\end{equation}
with 
\begin{equation}
C_0=-17.80\hspace{.3in}C_1=-0.002200
\end{equation}
from the automatic linear fitting (cf. Appendix Figure 2) whose coefficient
of determination $\mathbf{R}^{\mathbf{2}}=0.9951$.

The result of the two fittings can be summarized as follows: 
\begin{equation}
M_0\cong \langle M\rangle (z_0\rightarrow 0)=A_0\cong \langle M\rangle
(r\rightarrow 0)=C_0
\end{equation}

Of course $M_0$ represents the absolute magnitude of a hypothetical
supernova Ia with a central redshift $z_0\rightarrow 0$. As the Hubble
Magnitude $M$ is clearly an apparent absolute magnitude at increasing Hubble
depths, so its standard value for $D\rightarrow 0$ must necessarily coincide
with the true intrinsic absolute magnitude, that is $M_\alpha $ (cf. paper
XV).

The conclusion of the preliminary analysis of the SNe Ia absolute
magnitudes, based on the high-$z$ normal points with $\langle z\rangle >0.55$
of the paper IX Table 2, leads to the new result 
\begin{equation}
M_0=\langle M_\alpha \rangle (z_0\rightarrow 0)\cong -17.9
\end{equation}

The previous $M_0$ value agrees with the new absolute magnitudes $\langle
M^{*}\rangle $ of section 5, those listed in Table 4, that is with the
contents in the ADDENDUM NOTE - October 2011 - of paper XI.

\subsection{Construction of 30 new normal points from 398 $SCPU$ SNe data}

The normal points of paper IX Table 2 refer to excessively large $z$-ranges
to be able to represent accurately the SNe Hubble Magnitude trend at the low
redshifts of the nearby Universe. Therefore, in order to improve the
analysis, we need smaller $z$ bins. The following Table 6 and Table 7,
referring to the nearby and deep Universe respectively, collect 30 new
normal points, based on 398 $SCPU$ supernovae. These two tables were
constructed according to the same procedure as paper IX Table 2. In
particular the first 5 columns both of Table 6 and Table 7 contain numerical
values derived from the observed $z$ and $m_B^{\max }$ listed within the $%
SCPU$ compilation (Kowalski et al. 2008); the values referring to each $z$
bin are in the order: $z$ range; number N of the SNe included in the normal
point; unweighted mathematical mean $\langle m\rangle $ of the corresponding
SNe magnitudes $m_B^{\max }$; mean Hubble Magnitude $\langle M\rangle $
resulting from the normal eq. (2) applied to the bin, with $H_0=70$ assumed;
mathematical mean of the observed redshifts of the $z$ bin, according to the
position $\langle z\rangle \equiv z_0$ of eq. (1). The $6^{th}$ column holds
the value of the \textbf{Hubble Magnitude of a supernova Ia}, with $z$ $%
=\langle z\rangle \equiv z_0$ and $m=\langle m\rangle \equiv $ $m_0$ assumed
(cf. paper XV), according to the paper IX formula (19) (also called ECM $%
M(z_0)$ equation): 
\begin{equation}
M(z_0)=m_0-5\log \left[ cz_0\cdot (1+z_0)\right] +5\log H_0-25
\end{equation}

Fitting the points $M(z_0)$ plotted versus $z_0$ or $r(z_0)$ leads to the 
\textbf{line equation, }$M(z_0)$\textbf{\ or }$M(r)$\textbf{, representing
the central Hubble Magnitude of the supernovae Ia.}

The last two columns, $7^{th}$ and $8^{th}$, include two other \textbf{%
central quantities}, the light space $r(z_0)$ and the new absolute magnitude 
$M^{*}(z_0)$, corresponding to the assumed central redshift $z_0\equiv
\langle z\rangle $ and the central magnitude $m_0\equiv \langle m\rangle $.
Let us recall the ECM calculation procedure of $r(z_0)$, that applied in
section 2.1 of paper IX and section 4 of paper XV: 
\begin{equation}
z_0=\frac x3\left( \frac{1+x}{1-x}\right) \Rightarrow x=x(z_0)=\frac{%
3H_0r(z_0)}c\Rightarrow r(z_0)=\frac{cx(z_0)}{3H_0}
\end{equation}

According to eq. (31), the previous $r(z_0)$, whose values are listed in
column $7^{th}$ of Table 6 and 7, allow the introduction of a \textbf{new
central luminosity distance}, that is 
\begin{equation}
D_L^{*}(z_0)=r(z_0)\cdot (1+z_0)^2
\end{equation}
together with the \textbf{new absolute magnitude of a supernova Ia}, always
with $z$ $=\langle z\rangle \equiv $ $z_0$ and $m=\langle m\rangle \equiv $ $%
m_0$ assumed, as follows: 
\begin{equation}
M^{*}(z_0)=m_0-5\log \left[ r(z_0)\cdot (1+z_0)^2\right] -25
\end{equation}

Fitting the points $M^{*}(z_0)$ plotted versus $z_0$ or $r(z_0)$ leads to
the \textbf{line equation, }$M^{*}(z_0)$\textbf{\ or }$M^{*}(r)$\textbf{,
representing the new central absolute magnitude of the supernovae Ia}.

\newpage\ 

\textbf{Table 6}

\vspace{.07in}

\begin{tabular}{|l|l|l|l|l|l||l|l|}
\hline
$z$ range & N & $\langle m\rangle $ & $\langle M\rangle $ & $\langle
z\rangle \equiv z_0$ & $M(z_0)$ & $r(z_0)$ & $M^{*}(z_0)$ \\ \hline
$0<z\leq 0.010$ & $16$ & $14.24$ & $-18.14\pm 0.36$ & $0.007$ & $-18.24$ & $%
29$ & $-18.09$ \\ \hline
$0<z\leq 0.015$ & $33$ & $14.60$ & $-18.44\pm 0.21$ & $0.010$ & $-18.58$ & $%
40$ & $-18.48$ \\ \hline
$0.005\leq z\leq 0.020$ & $50$ & $14.99$ & $-18.63\pm 0.11$ & $0.013$ & $%
-18.75$ & $52$ & $-18.64$ \\ \hline
$0.010\leq z\leq 0.025$ & $45$ & $15.46$ & $-18.75\pm 0.11$ & $0.016$ & $%
-18.81$ & $63$ & $-18.60$ \\ \hline
$0.015\leq z\leq 0.030$ & $40$ & $15.90$ & $-18.86\pm 0.10$ & $0.021$ & $%
-18.92$ & $80$ & $-18.72$ \\ \hline
$0.020\leq z\leq 0.050$ & $39$ & $16.63$ & $-19.07\pm 0.06$ & $0.032$ & $%
-19.14$ & $116$ & $-18.84$ \\ \hline
$0.025\leq z\leq 0.100$ & $42$ & $17.19$ & $-19.14\pm 0.05$ & $0.044$ & $%
-19.28$ & $152$ & $-18.91$ \\ \hline
$0.030\leq z\leq 0.150$ & $37$ & $17.77$ & $-19.15\pm 0.05$ & $0.059$ & $%
-19.38$ & $193$ & $-18.90$ \\ \hline
$0.035\leq z\leq 0.200$ & $39$ & $18.42$ & $-19.16\pm 0.05$ & $0.083$ & $%
-19.51$ & $250$ & $-18.91$ \\ \hline
$0.040\leq z\leq 0.250$ & $42$ & $19.50$ & $-19.09\pm 0.06$ & $0.129$ & $%
-19.48$ & $340$ & $-18.68$ \\ \hline
$0.045\leq z\leq 0.300$ & $52$ & $20.04$ & $-19.09\pm 0.06$ & $0.163$ & $%
-19.51$ & $395$ & $-18.60$ \\ \hline
$0.050\leq z\leq 0.350$ & $66$ & $20.81$ & $-19.15\pm 0.06$ & $0.220$ & $%
-19.49$ & $473$ & $-18.43$ \\ \hline
$0.10\leq z\leq 0.40$ & $74$ & $21.80$ & $-19.13\pm 0.06$ & $0.291$ & $%
-19.24 $ & $552$ & $-18.02$ \\ \hline
$0.15\leq z\leq 0.45$ & $100$ & $22.20$ & $-19.18\pm 0.05$ & $0.341$ & $%
-19.26$ & $598$ & $-17.96$ \\ \hline
$0.20\leq z\leq 0.50$ & $120$ & $22.51$ & $-19.21\pm 0.05$ & $0.382$ & $%
-19.26$ & $632$ & $-17.90$ \\ \hline
\end{tabular}

\newpage\ 

\textbf{Table 7}

\vspace{.07in}

\begin{tabular}{|l|l|l|l|l|l||l|l|}
\hline
$z$ range & N & $\langle m\rangle $ & $\langle M\rangle $ & $\langle
z\rangle \equiv z_0$ & $M(z_0)$ & $r(z_0)$ & $M^{*}(z_0)$ \\ \hline
$0.25\leq z\leq 0.55$ & $131$ & $22.72$ & $-19.26\pm 0.04$ & $0.419$ & $%
-19.31$ & $660$ & $-17.90$ \\ \hline
$0.30\leq z\leq 0.60$ & $142$ & $22.87$ & $-19.32\pm 0.04$ & $0.450$ & $%
-19.36$ & $682$ & $-17.91$ \\ \hline
$0.35\leq z\leq 0.65$ & $143$ & $23.10$ & $-19.37\pm 0.04$ & $0.495$ & $%
-19.40$ & $711$ & $-17.90$ \\ \hline
$0.40\leq z\leq 0.70$ & $138$ & $23.25$ & $-19.44\pm 0.04$ & $0.533$ & $%
-19.47$ & $733$ & $-17.93$ \\ \hline
$0.45\leq z\leq 0.75$ & $118$ & $23.43$ & $-19.48\pm 0.04$ & $0.574$ & $%
-19.51$ & $756$ & $-17.93$ \\ \hline
$0.50\leq z\leq 0.80$ & $98$ & $23.61$ & $-19.55\pm 0.04$ & $0.623$ & $%
-19.57 $ & $781$ & $-17.96$ \\ \hline
$0.55\leq z\leq 0.85$ & $91$ & $23.82$ & $-19.60\pm 0.04$ & $0.680$ & $%
-19.62 $ & $808$ & $-17.97$ \\ \hline
$0.60\leq z\leq 0.90$ & $79$ & $24.03$ & $-19.62\pm 0.04$ & $0.730$ & $%
-19.64 $ & $829$ & $-17.94$ \\ \hline
$0.65\leq z\leq 0.95$ & $68$ & $24.25$ & $-19.68\pm 0.04$ & $0.797$ & $%
-19.69 $ & $855$ & $-17.96$ \\ \hline
$0.70\leq z\leq 1.00$ & $62$ & $24.42$ & $-19.71\pm 0.04$ & $0.851$ & $%
-19.72 $ & $875$ & $-17.96$ \\ \hline
$0.75\leq z\leq 1.10$ & $60$ & $24.52$ & $-19.74\pm 0.04$ & $0.885$ & $%
-19.75 $ & $886$ & $-17.97$ \\ \hline
$0.80\leq z\leq 1.20$ & $56$ & $24.62$ & $-19.79\pm 0.05$ & $0.927$ & $%
-19.80 $ & $900$ & $-18.00$ \\ \hline
$0.85\leq z\leq 1.30$ & $44$ & $24.77$ & $-19.86\pm 0.06$ & $0.996$ & $%
-19.88 $ & $921$ & $-18.05$ \\ \hline
$z\geq 0.9$ & $43$ & $25.01$ & $-19.88\pm 0.06$ & $1.082$ & $-19.91$ & $944$
& $-18.05$ \\ \hline
$z\geq 0.95$ & $34$ & $25.13$ & $-19.89\pm 0.07$ & $1.123$ & $-19.92$ & $955$
& $-18.04$ \\ \hline
\end{tabular}

\vspace{.2in}

Formally eq. (50) of $M^{*}(z_0)$ (whose high-$z$ values are listed in
column $8^{th}$ of the above Table 7) is different from eq. (32) of $\langle
M^{*}\rangle $ (whose high-$z$ values are listed in column $8^{th}$ of Table
4), that is the \textbf{normal equation of the new absolute magnitude, }here
rewritten in eq. (51), \textbf{\ } 
\begin{equation}
\langle M^{*}\rangle =\langle m\rangle -5\langle \log \left[
r_z(1+z)^2\right] \rangle -25
\end{equation}
where $r_z$ is the light space resulting from the \textbf{ECM }$z$\textbf{\
equation }(cf. eq. (4) of paper IX).

Fitting the normal points $\langle M^{*}\rangle $ plotted versus $z_0$ or $%
\langle r_z\rangle $ leads to the\textbf{\ line equation, }$\langle
M^{*}\rangle (z_0)$\textbf{\ or }$\langle M^{*}\rangle (\langle r_z\rangle )$%
\textbf{, representing the new normal absolute magnitude of the supernovae
Ia.}

Numerically, we find a small difference between $\langle M^{*}\rangle $ and $%
M^{*}(z_0)$, about $0.03$ magnitudes on average at high $z$, that is 
\begin{equation}
\langle M^{*}\rangle (z_0)-M^{*}(z_0)\approx 0.03
\end{equation}
\textbf{Thus the usefulness of the new central absolute magnitude }$%
M^{*}(z_0)$\textbf{\ is confirmed}.

\subsection{Plotting 30 values of SNe $\langle M\rangle $, $M(z_0)$, $%
M^{*}(z_0)$ versus $z_0$ and $r(z_0)$}

The 30 values of $\langle M\rangle $, $M(z_0)$, $M^{*}(z_0)$ in Table 6 and
7 from $SCPU$ data of 398 SNe allow the construction of the corresponding 6
plots, versus $z_0\equiv \langle z\rangle $ and $r(z_0)$ respectively. These
diagrams appear in the Appendix ''Atlas of the ECM paper XVI figures''. In
particular Appendix Figure 16 presents the plot of 30 SNe Ia normal Hubble
Magnitudes $\langle M\rangle $ versus the mean redshift $\langle z\rangle $,
Appendix Figure 17 the plot of 30 SNe Ia normal Hubble Magnitudes $\langle
M\rangle $ versus the ECM $r(z_0)$, Appendix Figure 18 the plot of 30 SNe Ia
central Hubble Magnitudes $M(z_0)$ versus $z_0$, Appendix Figure 19 the plot
of 30 SNe Ia central Hubble Magnitudes $M(z_0)$ versus the ECM $r(z_0)$,
Appendix Figure 20 the plot of 30 SNe Ia central absolute magnitudes $%
M^{*}(z_0)$ versus $z_0$, Appendix Figure 21 the plot of 30 SNe Ia central
absolute magnitudes $M^{*}(z_0)$ versus the ECM $r(z_0)$.

\subsection{The magnitude anomaly of the SNe Ia at low $\langle z\rangle $}

Even at first sight the plots of the Appendix Figures 16-17-18-19-20-21
highlight the magnitude anomaly of the low $\langle z\rangle $\ points. In
other words these six diagrams give clear empirical evidence for the normal
luminosity behaviour of the supernovae Ia of the deep Universe in comparison
with the SNe magnitude trend of the nearby Universe. Such a distinction has
been emphasized through the separation of the 30 normal points into two
groups of 15 points each. Table 6 collects 15 normalized-central supernovae
Ia, which appear to be affected by the magnitude anomaly, with individual
redshifts $z\leq 0.5$, while Table 7 collects other 15 normalized-central
supernovae Ia based on individual redshifts $z\geq 0.25$. In particular it
is remarkable to see in Appendix Figure 20 a significant linear trend
(almost constant) of the central absolute magnitudes $M^{*}(z_0)$ after high
normal redshifts, with $\langle z\rangle \gtrsim 0.4$. Thus a preliminary
cut-off redshift limit between the nearby Universe affected by the magnitude
anomaly and the unperturbed deep Universe is here fixed at $z=0.25$ and
corresponding $\langle z\rangle >0.4$. But the discovered variation of the
SNe Ia luminosity may be only apparent, because there is no astrophysical
explanation able to reproduce intrinsically the observed maximum peak in the
depth range $0.04\lesssim $\ $\langle z\rangle \lesssim 0.08$, with a
resulting $\Delta M\approx 1$ (cf. Appendix Figures 16-18-20).

\subsection{Astronomical evidence for cosmic rotation}

An interpretation of the observed magnitude anomaly can be found in paper
VII ''Cosmic mechanics of the nearby Universe within the expansion center
model with angular momentum conserved''. In other words the negative
collapse of the SNe $M$ at $\langle z\rangle \approx 0.06$ and $\langle
z\rangle \equiv z_0\lesssim 0.4$ is here considered to be a proof of cosmic
rotation, which not even the ECM Hubble law (cf. section 2.1 and papers
V-VI-IX) includes. Consequently the related magnitude formula, owing to the
inclusion of distorted Hubble depths $D=cz_0/H_0=cz/H_X$ or light spaces $r$
as inferred from the ECM Hubble law, should also give distorted values of
SNe $\langle M\rangle $, $M(z_0)$, $M^{*}(z_0)$ to a wide Galaxy entourage,
including the Huge Void (Bahcall \& Soneira 1982) and the expansion center
at $R_0\approx 260$ $Mpc$ from the Local Group (cf. papers I-II and author
1991). Indeed, only the very nearby Universe, at $z_0\lesssim 0.007$ or $%
D\lesssim 30$ $Mpc$, should be somewhat independent from the cosmic
rotation, owing to the Galilean relativity effect within the ECM rigid
rotation; on the other hand also the normal or central points of the deep
Universe, at $\langle z\rangle \equiv z_0\gtrsim 0.4$ or $D\gtrsim 1000$ $%
Mpc $, result to be negligibly affected by the cosmic rotation, probably
thanks to a better statistical merging of the individual $z$ points. Here we
must remark that, according to the rotating Universe calculated in paper
VII, the transversal velocity of the Galaxy, $R_0\dot \vartheta _0\approx
6\times 10^9cm/s$, is more than three times the radial velocity, $\dot
R_0\approx 1.8\times 10^9cm/s$. Therefore the observed redshift $z$ from the
Milky Way must also be linked to a relative motion of differential rotation,
which however is inconsistent with the ECM rigid rotation. In conclusion the
magnitude anomaly of the SNe Ia at low $\langle z\rangle $ may be
technically interpreted as due to a deficiency in the used magnitude
formulas, which produce a maximum peak of deviation, with a resulting
systematic $\Delta M\approx 1$ at $0.04\lesssim \langle z\rangle \lesssim
0.08$, that is in the Hubble depth range $170$ $Mpc\lesssim $\ $D\lesssim
350 $ $Mpc$.

\subsection{Fitting 15 values of High-$z$ SNe $\langle M\rangle $, $M(z_0)$, 
$M^{*}(z_0)$ versus $z_0$ and $r$}

As a consequence of the previous results, a correct analysis of the SNe Ia
absolute magnitudes (cf. eqs. 39-42-47-50) must necessarily be limited to
the data of Table 7, that of a deep Universe whose magnitude anomaly seems
to be negligible within the limits of the present astronomical measurements.

Fitting the 15 points $\langle M\rangle $ (cf. Table 7) plotted versus $z_0$
and $r(z_0)$ leads to the line equations, $\langle M\rangle (z_0)$\ and $%
\langle M\rangle (r)$, representing the normal Hubble Magnitude of the
supernovae Ia,\ as a function of the central redshift $z_0$ and light space $%
r(z_0)$. The solutions from the following automatic cubic and linear
fittings (cf. Appendix Figures 22-23) 
\begin{equation}
\langle M\rangle (z_0)=A_0+A_1z_0+A_2z_0^2+A_3z_0^3
\end{equation}
\begin{equation}
\langle M\rangle (r)=C_0+C_1r(z_0)
\end{equation}
, whose corresponding coefficients of determination $\mathbf{R}^{\mathbf{2}%
}=0.9948$ and $\mathbf{R}^{\mathbf{2}}=0.9950$ respectively, give the
values: 
\begin{equation}
A_0=-18.26\hspace{.2in}A_1=-3.351\hspace{.2in}A_2=+2.636\hspace{.2in}%
A_3=-0.8485
\end{equation}
\begin{equation}
C_0=-17.86\hspace{.3in}C_1=-0.002138
\end{equation}

Fitting the 15 points $M(z_0)$, (cf. Table 7) plotted versus $z_0$ and $%
r(z_0)$ leads to the line equations, $M(z_0)$\ and $M(r)$, representing the
central Hubble Magnitude of the supernovae Ia,\ as a function of the central
redshift $z_0$ and light space $r(z_0)$. The solutions from the following
automatic cubic and linear fittings (cf. Appendix Figures 24-25) 
\begin{equation}
M(z_0)=A_0+A_1z_0+A_2z_0^2+A_3z_0^3
\end{equation}
\begin{equation}
M(r)=C_0+C_1r(z_0)
\end{equation}
, whose corresponding coefficients of determination $\mathbf{R}^{\mathbf{2}%
}=0.9939$ and $\mathbf{R}^{\mathbf{2}}=0.9923$ respectively, give the
values: 
\begin{equation}
A_0=-18.38\hspace{.2in}A_1=-3.188\hspace{.2in}A_2=+2.681\hspace{.2in}%
A_3=-0.9603
\end{equation}
\begin{equation}
C_0=-17.95\hspace{.3in}C_1=-0.002061
\end{equation}

Fitting the 15 points $M^{*}(z_0)$(cf. Table 7) plotted versus $z_0$ and $%
r(z_0)$ leads to the line equations, one of $M^{*}(z_0)$\ and two of $%
M^{*}(r)$, representing the new central absolute magnitude of the supernovae
Ia,\ as a function of the central redshift $z_0$ and light space $r(z_0)$.
The solutions from one quadratic and two linear automatic fittings (cf.
Appendix Figures 26-27-28) 
\begin{equation}
M^{*}(z_0)=A_0+A_1z_0+A_2z_0^2
\end{equation}
\begin{equation}
M^{*}(z_0)=A_0+A_1z_0
\end{equation}
\begin{equation}
M^{*}(r)=C_0+C_1r(z_0)
\end{equation}
, whose corresponding coefficients of determination $\mathbf{R}^{\mathbf{2}%
}=0.8883$, $\mathbf{R}^{\mathbf{2}}=0.8772$ and $\mathbf{R}^{\mathbf{2}%
}=0.8353$, respectively, give the values: 
\begin{equation}
A_0=-17.87\hspace{.2in}A_1=-0.02420\hspace{.2in}A_2=-0.1199
\end{equation}
\begin{equation}
A_0=-17.81\hspace{.2in}A_1=-0.2071
\end{equation}
\begin{equation}
C_0=-17.57\hspace{.2in}C_1=-4.831E-04
\end{equation}

\subsection{A final solution for the SNe Ia $\langle M\rangle $ and $\langle
M^{*}\rangle $}

Each solution in the previous sections has given an extrapolated value of
the absolute magnitude $\langle M_\alpha \rangle (z_0\rightarrow 0)$ as $%
M_0= $ $A_0$ or $M_0=$ $C_0$. Thus the solution here adopted for the
absolute magnitude\ $M_0$ of the supernovae Ia, from the mathematical mean
of the $9$ values listed above, is the following: 
\begin{equation}
M_0=-17.9\pm 0.1
\end{equation}

Once the starting point has been fixed at this $M_0=-17.9$, a solution of
the SNe Ia $\langle M\rangle $ and $\langle M^{*}\rangle $ can be found
taking into account \textbf{only the best normal points, that is the core of
the available data}, those based on $z$ bins with individual $z\geq $ $0.4$
and a number $N\geq 60$ as a minimum limit for the SNe included in the
normal point. In particular the choice for $\langle M\rangle $ includes $9$
normal points from paper IX Table 2 and the $4$ ECM normal points here
listed as $\langle M_z\rangle $ in Table 8, while that for $\langle
M^{*}\rangle $ includes only the $4$ ECM normal points of Table 8.

Fitting the plot of the $9$ core points $\langle M\rangle $ from paper IX
Table 2 and the starting point $M_0=-17.9$, both versus $\langle z\rangle
\equiv z_0$ and $r(z_0)$, leads to the line equations, $\langle M\rangle
(z_0)$\ and $\langle M\rangle (r)$, representing \textbf{the normal Hubble
Magnitude of the supernovae Ia},\ as a function of the central redshift $z_0$
and light space $r(z_0)$. The solutions of the following automatic cubic and
linear fittings (cf. Appendix Figures 29-31) 
\begin{equation}
\langle M\rangle (z_0)=A_0+A_1z_0+A_2z_0^2+A_3z_0^3
\end{equation}
\begin{equation}
\langle M\rangle (r)=C_0+C_1r(z_0)
\end{equation}
, whose corresponding coefficients of determination $\mathbf{R}^{\mathbf{2}%
}=0.99992$ and $\mathbf{R}^{\mathbf{2}}=0.9996$ respectively, give the
values: 
\begin{equation}
A_0=-17.900\hspace{.2in}A_1=-4.2618\hspace{.2in}A_2=+3.2507\hspace{.2in}%
A_3=-0.90878
\end{equation}
\begin{equation}
C_0=-17.90\hspace{.3in}C_1=-0.002091
\end{equation}
The parallel check solutions based only on the $9$ points $\langle M\rangle $
without $M_0=-17.9$ (cf. Appendix Figures 30-32) give $\mathbf{R}^{\mathbf{2}%
}=0.9974$ and $\mathbf{R}^{\mathbf{2}}=0.9943$, with $A_0=-18.11$ and $%
C_0=-17.75$, respectively.

An alternative solution is based on the $4$ ECM normal points of Table 8,
which was constructed by combining the core points from Table 0 of section
3.1 with those from Table 4 of section 5.

\newpage\ 

\textbf{Table 8}

\vspace{.1in}

\begin{tabular}{|c|c|c|c|c|c|c|}
\hline
N & $z$ bin & $\langle z\rangle $ & $\langle m\rangle $ & $\langle
r_z\rangle $ & $\langle M_z\rangle $ & $\langle M^{*}\rangle $ \\ \hline
200 & $0.4\leq z<1.4$ & $0.677$ & $23.763$ & $793.0$ & $-19.567$ & $-17.928$
\\ \hline
149 & $0.5\leq z<1.4$ & $0.756$ & $24.052$ & $830.4$ & $-19.650$ & $-17.951$
\\ \hline
107 & $0.6\leq z<1.4$ & $0.837$ & $24.339$ & $865.6$ & $-19.719$ & $-17.961$
\\ \hline
75 & $0.7\leq z<1.4$ & $0.919$ & $24.627$ & $898.9$ & $-19.773$ & $-17.954$
\\ \hline
\end{tabular}

\vspace{.2in}

Fitting the plot of the $4$ ECM points $\langle M_z\rangle $ of Table 8 and
the starting point $M_0=-17.9$, both versus $\langle z\rangle \equiv z_0$
and $\langle r_z\rangle $, leads to the line equations, $\langle M_z\rangle
(z_0)$\ and $\langle M_z\rangle (\langle r_z\rangle )$, representing \textbf{%
the ECM normal Hubble Magnitude of the supernovae Ia},\ as a function of the
central redshift $z_0$ and light space $\langle r_z\rangle $. The solutions
of the following automatic cubic and linear fittings (cf. Appendix Figures
33-35) 
\begin{equation}
\langle M_z\rangle (z_0)=A_0+A_1z_0+A_2z_0^2+A_3z_0^3
\end{equation}
\begin{equation}
\langle M_z\rangle (\langle r_z\rangle )=C_0+C_1\langle r_z\rangle
\end{equation}
, whose corresponding coefficients of determination $\mathbf{R}^{\mathbf{2}%
}=1.0000$ and $\mathbf{R}^{\mathbf{2}}=0.99990$ respectively, give the
values: 
\begin{equation}
A_0=-17.900\hspace{.2in}A_1=-4.0675\hspace{.2in}A_2=+2.8270\hspace{.2in}%
A_3=-0.67334
\end{equation}
\begin{equation}
C_0=-17.901\hspace{.3in}C_1=-0.0020968
\end{equation}
The parallel check solutions based only on the $4$ points $\langle
M_z\rangle $ without $M_0=-17.9$ (cf. Appendix Figures 34-36) give $\mathbf{R%
}^{\mathbf{2}}=0.9999$ and $\mathbf{R}^{\mathbf{2}}=0.9954$, with $%
A_0=-18.12 $ and $C_0=-18.03$, respectively.

Indeed, the high reliability of the $4$ ECM normal points of Table 8 is
clearly shown by the very precise solutions above listed, which are very
near to those derived from the previous $9$ points $\langle M\rangle $ from
paper IX Table 2. Consequently these $4$\textbf{\ ECM normal points are}
here considered \textbf{pilot points} also for finding a better trend of the
new absolute magnitude $M^{*}$ of the supernovae Ia.

Fitting only the plot of the $4$ core points $\langle M^{*}\rangle $ listed
in Table 8 (excluding the starting point $M_0=-17.9$), both versus $\langle
z\rangle \equiv z_0$ and $\langle r_z\rangle $, leads to the line equations, 
$\langle M^{*}\rangle (z_0)$ and $\langle M^{*}\rangle (\langle r_z\rangle )$%
, representing \textbf{the new normal absolute magnitude of the supernovae Ia%
},\ as a function of the central redshift $z_0$ and light space $\langle
r_z\rangle $, with $\langle M^{*}\rangle (z_0)\equiv \langle M^{*}\rangle
(\langle r_z\rangle )\equiv \langle M_\alpha \rangle (z_0)$ assumed. The
solutions from both the automatic linear fittings (cf. Appendix Figures
37-38), that is 
\begin{equation}
\langle M^{*}\rangle (z_0)=A_0+A_1z_0
\end{equation}
and 
\begin{equation}
\langle M^{*}\rangle (\langle r_z\rangle )=C_0+C_1\langle r_z\rangle
\end{equation}
, whose corresponding coefficients of determination $\mathbf{R}^{\mathbf{2}%
}=0.6237$ and $\mathbf{R}^{\mathbf{2}}=0.6564$ respectively, give the
values: 
\begin{equation}
A_0=-17.86\hspace{.2in}A_1=-0.1084
\end{equation}
\begin{equation}
C_0=-17.73\hspace{.2in}C_1=-0.0002541
\end{equation}

The $6$ previous fittings carried out without the starting point $M_0=-17.9$
give again an extrapolated absolute magnitude $\langle M_\alpha \rangle
(z_0\rightarrow 0)$ as $M_0=$ $A_0$ or $M_0=$ $C_0$. Hence the computed
solution for the absolute magnitude\ $M_0$ of the supernovae Ia, from the
mathematical mean of the $6$ values listed above, is here confirmed to be
the following: 
\begin{equation}
M_0=-17.93\pm 0.08
\end{equation}

Finally, the solutions here proposed for the SNe $\langle M\rangle $ and $%
\langle M^{*}\rangle $ permit the computation of both the total $M$ spread
and the absolute magnitude $M_\alpha $ when $M_\alpha \equiv M^{*}$ is
assumed, according to paper XV and paper X Appendix. Table 9 lists $5$
spread values (in second, fourth and sixth column) following the $3$
solutions (70)(74)(78), calculated at the $5$ different $\langle z\rangle
\equiv z_0$ of the first column. In addition Table 9 also reports the
relativistic value of the deceleration parameter which results by applying
the total spread of the extrapolated Hubble Magnitudes $\langle M\rangle
(z_0)$ and $\langle M_z\rangle (z_0)$ at $z_0=0.001$ into the $q_0$ formula
(59) of the parallel paper XV (or A19 of paper X).

\vspace{.1in}

\textbf{Table 9}

\vspace{.1in}

\begin{tabular}{|c|c|c|c|c|c|}
\hline
\multicolumn{1}{|c||}{$z_0$} & $\langle M\rangle (z_0)-M_0$ & 
\multicolumn{1}{|c||}{$q_0$} & $\langle M_z\rangle (z_0)-M_0$ & 
\multicolumn{1}{|c||}{$q_0$} & $\langle M^{*}\rangle (z_0)-A_0$ \\ \hline
\multicolumn{1}{|c||}{0.001} & $-0.004259$ & \multicolumn{1}{|c||}{$+2.92$}
& $-0.004065$ & \multicolumn{1}{|c||}{$+2.74$} & $-0.000109$ \\ \hline
\multicolumn{1}{|c||}{0.01} & $-0.04229$ & \multicolumn{1}{|c||}{} & $%
-0.04039$ & \multicolumn{1}{|c||}{} & $-0.00108$ \\ \hline
\multicolumn{1}{|c||}{0.1} & $-0.3946$ & \multicolumn{1}{|c||}{} & $-0.3792$
& \multicolumn{1}{|c||}{} & $-0.0108$ \\ \hline
\multicolumn{1}{|c||}{0.5} & $-1.432$ & \multicolumn{1}{|c||}{} & $-1.411$ & 
\multicolumn{1}{|c||}{} & $-0.0542$ \\ \hline
\multicolumn{1}{|c||}{1} & $-1.920$ & \multicolumn{1}{|c||}{} & $-1.914$ & 
\multicolumn{1}{|c||}{} & $-0.1084$ \\ \hline
\end{tabular}

\vspace{.1in}

\subsection{The new absolute magnitude $M^{*}$}

At the end of this magnitude analysis, the coincidence between the intrinsic
absolute magnitude $M_\alpha $ with the new absolute magnitude $M^{*}$ (cf.
paper XV, paper XI Addendum Note, paper X Appendix) must also be shown
theoretically, summed up in the identity 
\begin{equation}
M_\alpha \equiv M^{*}=m-5\log D_L-25
\end{equation}
with 
\begin{equation}
D_L=r\cdot (1+z)^2=r_0\cdot (1+z)
\end{equation}

as a new formulation of the luminosity distance $D_L$, which differs from
relativistic cosmology in that, here, the light space $r=-c\Delta t$ is a
physical distance, representing the space run by light during the past
travel time $\Delta t=t-t_0$, in place of the relativistic proper distance $%
r_{pr}$ at the emission epoch $t$ (cf. section 2 of paper VIII, IX, XV).

Mathematically, such light-space $r$ in eq. (82) is the same $r$ we find in
Milne's cosmology (Rowan-Robinson 1996) as the distance at the emission
epoch; however the ''cosmic medium'' (CM), with respect to which light moves
at constant speed $c=\lambda /T$, is expanding as does the whole Universe.
Consequently, also $\lambda $ and $T$ increase, because of the CM expansion
with the constancy of $c$. As a result, the light-space $r$ is larger than
the distance at the emission epoch, although its value in light-time
represents a measure of that past epoch $t$.

The same, $r_0=r\cdot (1+z)$ in eq. (82) seems to substitute the
relativistic proper distance at the present epoch $t_0$, while its meaning
has yet to be found within expansion center cosmology.

In conclusion the physical demonstration of eq. (82) is possible, but such a
task must belong rigorously to the theoreticians.

\newpage 

\section{Conclusions}

The present paper, after the parallel paper XV, is the latest in a series of
ECM papers, containing important new results, based on the fundamental SCP
Union data, in addition to the further reconfirmation of the expansion
center model.

A few remarks and concluding statements on the topic may be summarized as
follows:

0) The preliminary empirical model of the expansion center Universe (Lorenzi
1989) led to a confirmed dipole equation of the Hubble ratio (Lorenzi
1991-93); after the discovery of a high rate of variation of the Hubble
constant in the nearby Universe (Lorenzi 1994), a more rigorous formulation
of the new Hubble law was developed and studied in terms of possible
outcomes. The expansion center model (ECM) is the proposed solution in the
1999 papers I and II;

1) The adjective ''model independent'', first of all means the exclusion of
the ECM formalism of paper II, as specified in the ''Introduction'' of paper
XV. In this sense papers V, VI, IX, XV may be considered model independent
papers;

2) The formulation for the wedge-shape of the new Hubble law, or new Hubble
D law, implies a clear Hubble ratio dipole, as $cz/D=H_0-a^{*}(D)\cos \gamma 
$, where the Hubble depth $D$ $=D_C/(1+z)$ can be calculated through the $M$
value which produces the identity $D_C\equiv $ $D_L$ $=10^{0.2(m-M)-5}$;

3) The relation (22) in paper IX, expressing the average trend of the SNe
Hubble Magnitude $M(z_0)$ $=d_0+d_1D+d_2D^2$ $\equiv \langle M\rangle $, was
constructed by using the normal ECM $M$ equation (21) from paper IX, with $%
H_0=70$ H.u. assumed and without including the ECM dipole terms of 398 $SCPU$
supernovae. In practice that means the adoption of an ECM-independent
procedure;

4) Without doubt, paper XV has produced two significant results, that is a
model independent confirmation both of the Hubble ratio dipole and of the
angular coefficient $a^{*}=$ $5.5$ H.u. predicted by the ECM at the central
redshift $z_0\equiv $ $\langle z\rangle =1.0$ or Hubble depth $D\cong 4283$ $%
Mpc$;

5) Unlike in paper XV, the dipole analysis of this paper XVI was based on
the adoption of the ECM, to make possible the check test of the ECM standard
value, $a_0=$ $12.7$ H.u.. A similar procedure was applied also in paper VI
and in its integral version;

6) A new finding from paper XVI is a clear macroscopic discovery of large $%
\Delta M$ in SNe Ia, which consequently, at the present time, are not usable
standard candles when taken individually;

7) A secondary result from paper XVI is the resulting drop in the scattering 
$\Delta M$ with the Hubble depth $D$, more likely according to the
relationship $\langle \left| \Delta M\right| \rangle \cong 1.4-0.15\ln (D)$;

8) In addition to the previous relationship, the available data set of Table
3 seems to suggest the preliminar correlation $\langle \left| \Delta
M\right| \rangle \approx 0.34-0.10\cdot \cos \gamma $ , only at redshifts $%
z\lesssim 0.5$ (cf. Appendix Figure 7);

9) Another important outcome of the present ''Dipole analysis...'' is the
evidence for a clear perturbation effect on ECM of the SNe $\Delta M$ at $%
z\lesssim 0.5$ ;

10) The above cited perturbation effect of $\Delta M$, after introducing the
weights $w_i\propto \left| \Delta M\right| ^{-i}$, allows both a further
reconfirmation of the expansion center model at any Hubble depth $D$ and, at
the same time, the demonstration of the adopted $\langle M\rangle $ $%
=d_0+d_1D_z+d_2D_z^2$ being able to accurately reproduce the predicted
Hubble ratio dipole when $\Delta M\rightarrow 0$ ;

11) The unweighted dipole tests, that is with $w_0=1$, directly confirm the
ECM at $z\gtrsim 0.5$, since the mathematical mean of all the 10 $a_0$
values, those resulting from W18-W19-W20-W21-W22 of Table 1 and
A18-A19-A20-A21-A22 of Table 2, becomes $\langle a_0\rangle =12.0\pm 0.6$
H.u., while the above best 5 fittings in Table 1 give $\langle a_0\rangle
=12.8\pm 0.7$ H.u.. Once again these average values agree very well with the
ECM, being $12.7$ H.u the standard value of $a_0$;

12) A further result here reported is some astronomical evidence for
intrinsic SNe Ia $\Delta M$ ;

13) The magnitude anomaly of the $SCPU$ supernovae at low redshifts, with an
observed maximum peak of $\Delta M\approx 1$ in the range $0.04\lesssim $\ $%
\langle z\rangle \lesssim 0.08$ (cf. Appendix Figures 16-18-20), is the most
important finding in paper XIII, which has been rolled out in paper XVI ;

14) The negative collapse of the SNe $M$ at $\langle z\rangle \approx 0.06$
in a range $0.007\lesssim $ $\langle z\rangle \lesssim 0.4$ is here
considered to be structural and due to the cosmic rotation, which should
affect significantly the usual magnitude formulas for a wide Galaxy
entourage, including the Huge Void (Bahcall \& Soneira 1982) and the
expansion center at $R_0\approx 260$ $Mpc$ (cf. ECM papers I-II and author
1991);

15) Once the perturbation zone on the SNe $M$ is removed, the luminosity
analysis of high $z$ SNe Ia has allowed the extrapolation of the
corresponding absolute magnitude $M_0$ value at a central redshift $%
z_0\rightarrow 0$. The final result is $M_0=-17.93\pm 0.08$ ;

16) The extrapolated trend of the normal Hubble Magnitude $\langle M\rangle $
of the supernovae Ia at low central redshifts $z_0\equiv \langle z\rangle
\ll 1$, according to $\langle M\rangle =\langle m\rangle -5\langle \log
\left[ D(1+z)\right] \rangle -25$ with $D=cz/H_X\equiv cz_0/H_0$, presents a
sharp negative increase with $z_0$, which clearly contrasts with the almost
constant trend due to a relativistic $q_0\approx -1$ (cf. paper XV and paper
X Appendix);

17) The new ECM absolute magnitude of the supernovae Ia, that $M^{*}$ based
on a luminosity distance $D_L=r_z\cdot (1+z)^2$ where $r_z=-c(t-t_0)$ is the
light space resulting from the \textbf{ECM }$z$\textbf{\ equation} as space
run by light at constant speed $c$ into the expanding ''cosmic medium'' or
Hubble flow, shows here a slowly increasing negative trend, that is: $%
\langle M^{*}\rangle =-17.9-0.1\times z_0$, with $z_0\equiv \langle z\rangle 
$ assumed;

18) Two precise values of the determination coefficient, that is $\mathbf{R}%
^{\mathbf{2}}=0.99992$ and $\mathbf{R}^{\mathbf{2}}=1.0000$, from the final
cubic fittings of $\langle M\rangle (z_0)$ and $\langle M_z\rangle (z_0)$
respectively, give the corresponding total $M$ spread in Table 9 a high
accuracy. As a consequence, the more reliable value of the relativistic
deceleration parameter $q_0$ here is about $+3$ ;

19) The intrinsic absolute magnitude $M_\alpha $ is found to coincide with
the new absolute magnitude $M^{*}$, that is $M_\alpha \equiv M^{*}$, based
both on empirical and theoretical results;

20) After the strong experimental evidence for the expansion center and some
mechanical investigations about the Universe as a whole, according to the
ECM papers series, this paper XVI presents a noteworthy observational proof
of the cosmic rotation, that is the magnitude anomaly of the nearby
supernovae Ia. Thus Gamow (1946) was right to propose a ''Rotating
Universe?'' to Einstein, however unsuccessfully (cf. Kragh 1996). Actually
there are other important astronomical proofs on the topic (cf. Longo 2011).
The conclusion might be in favour of a Big Bang as a Big Crush, when the ECM
cosmic mechanics with angular momentum conserved (cf. paper VII and VIII) is
applied even to Lema\^\i tre primitive atom (1946).

\newpage\ 

\section{Acknoledgements}

The present analysis has been made possible thanks to the SCP team and their
Union compilation. The author would like to thank Ms. Laura Daricello of the
Astronomical Observatory of Palermo and the Local Organizing Committee of
the successful meeting SAIt2011 for their kind support. Finally the author
wishes to express his sincere gratitude to the Italian Astronomical Society
and the President Roberto Buonanno, for the constant official and scientific
attention given to the research on the expansion center Universe.

\newpage

\section{References}

\vspace{0.2in}

\hspace{0.2in}Astier, P. et al. 2006, A\&A 447, 31

Bahcall, N.A. \& Soneira, R.M. 1982, ApJ 262, 419 (B\&S)

EWASS 2012, http://www.ifsi-roma.inaf.it/ewass2012/

Gamow, G. 1946, ''Rotating Universe?'', Nature, 158, 549

Harvard-IAU 2003, http://cfa-www.harvard.edu/iau/lists/Supernovae.html

Knop, R.A. et al. 2003, ApJ 598, 102 (K03)

Kowalski, M. et al. 2008, arXiv:0804.4142v1 $\left[ \text{astro-ph}\right] $
25 Apr 2008$\rightarrow $ApJ 686, 749

Kragh, H. 1996, ''Cosmology and Controversy'', Princeton University Press

Lema\^\i tre, G. 1946, ''L'Hypothese de l'atom primitif'', Neuchatel, Griffon

Longo, M.J. 2011, arXiv:1104.2815v1 $\left[ \text{astro-ph.CO}\right] $ 14
Apr 2011

Lorenzi, L. 1989, 1991, Contributi N. 0,1, Centro Studi Astronomia -
Mondov\`\i , Italy

\hspace{.5in}1993, in 1995 MemSAIt, 66, 1, 249

\hspace{.5in}1994, in 1996 Astro. Lett. \& Comm., 33, 143

\hspace{.5in}1999a, astro-ph/9906290 17 Jun 1999,

\hspace{.5in}in 2000 MemSAIt, 71, 1163 (paper I: reprinted in 2003, MemSAIt,
74)

\hspace{.5in}1999b, astro-ph/9906292 17 Jun 1999,

\hspace{.5in}in 2000 MemSAIt, 71, 1183 (paper II: reprinted in 2003,
MemSAIt, 74)

\hspace{.5in}2003b, MemSAIt Suppl. 3,

\hspace{.5in}http://sait.oat.ts.astro.it/MSAIS/3/POST/Lorenzi\_poster.pdf
(paper V)

\hspace{.5in}2004, MemSAIt Suppl. 5, 347 (paper VI: partial and integral
version)

\hspace{.5in}2008,
http://terri1.oa-teramo.inaf.it/sait08/slides/I/ecmcm9b.pdf (paper VII)

\hspace{.5in}2009,
http://astro.df.unipi.it/sait09/presentazioni/AulaMagna/08AM/lorenzi.pdf

\hspace{.5in}(paper VIII)

\hspace{.5in}2010, arXiv: 1006.2112v3 $\left[ \text{physics.gen-ph}\right] $
17 Jun 2010 (paper IX)

\hspace{.5in}2011a,
http://www.astropa.unipa.it/SAIT2011/Proceedings/Lorenzi1.pdf (paper X)

\hspace{.5in}2011b,
http://www.astropa.unipa.it/SAIT2011/Proceedings/Lorenzi2.pdf (paper XI)

\hspace{.5in}2012a, poster paper presented at EWASS 2012 (paper XII)

\hspace{.5in}2012b, parallel poster paper presented at EWASS 2012 (paper
XIII)

\hspace{.5in}2012d (paper XV: parallel paper)

\newpage\ 

Miknaitis, G. 2007, ApJ 666, 674

Perlmutter, S., et al. 1999, ApJ 517, 565 (P99)

Riess, A.G. et al. 2004, ApJ 607, 665

Riess, A.G. et al. 2007, ApJ 659, 98

Rowan-Robinson, M. 1996, ''Cosmology'' Clarendon Press - Oxford

Sandage, A., Tammann G. A. 1975a, ApJ 196, 313 (S\&T: Paper V)

Wikipedia 2011, http://en.m.wikipedia.org/wiki/Allan\_Sandage

\newpage\ 

\section{APPENDIX: Atlas of the ECM paper XVI figures}

All the plots and graphical fittings of this ''Dipole and absolute magnitude
analysis of the SCP Union supernovae ...'' appear in the following check
atlas of 38 figures and their corresponding legends.

The atlas uses Hubble units; therefore the abscissae as Hubble depth $D_z$
or the mean $\langle D_z\rangle $, light space $r(z_0)$ or $\langle
r_z\rangle $ are in Megaparsecs, while $\langle z\rangle \equiv z_0$ is
normal redshift; the abscissae as $\cos \gamma $, $-X$ or the mean $\langle
-X\rangle $ are dimensionless; the ordinates as $\left| \Delta M\right| $, $%
\langle \left| \Delta M\right| \rangle $, $M_z$, $\langle M_z\rangle $, $%
\langle M\rangle $, $M(z_0)$, $M^{*}(z_0)$, $\langle M^{*}\rangle $ are
magnitudes; the ordinates as $Y$ are in $km$ $s^{-1}Mpc^{-1}$.

In the cartesian plane $(x,y)$

of Figures
3-4-5-6-8-9-10-11-12-13-14-15-22-23-24-25-26-27-28-29-30-31-32-33-34-35-36-37-38

the resulting fitting equations, as $y=f(x)$, are included, together with
the coefficient of determination $\mathbf{R}^2$.

The diagrams of Figures 16-17-18-19-20-21 highlight the magnitude anomaly of
the low $\langle z\rangle $\ points. In particular \textbf{Figure 20}, that
presents the plot of 30 SNe new central absolute magnitudes $M^{*}(z_0)$
versus $\langle z\rangle =z_0$ from SCP Union data of 398 supernovae Ia,
gives clear empirical evidence for the normal luminosity behaviour of the
supernovae Ia of the deep Universe in comparison with the SNe Ia magnitude
trend of the nearby Universe, where we can see a maximum peak of $M^{*}$
deviation, with a resulting systematic $\Delta M^{*}\approx 1$ at $%
0.04\lesssim \langle z\rangle \lesssim 0.08$, that is in the Hubble depth
range $170$ $Mpc\lesssim $\ $D\lesssim 350$ $Mpc$. Note that the distance of
the expansion center from the Local Group at the present epoch $t_0$ results
to be $R_0\approx 260$ $Mpc$, according to the ECM.

Lastly, the high reliability of the core points in Table 8 is clearly shown
by the plots and precise fittings of Figures 33-34-35-36. Thus these $4$%
\textbf{\ ECM normal points }become \textbf{pilot points }also in Figures
37-38, to represent two linear trends of the new normal absolute magnitude $%
\langle M^{*}\rangle $ of the supernovae Ia.

\end{document}